\documentclass{aa}  

\usepackage{siunitx}
\usepackage{graphicx}
\usepackage{txfonts}
\usepackage{natbib}
\usepackage{lscape}
\usepackage[
        colorlinks=true,
        citecolor=blue,
        linkcolor=blue]{hyperref}

\newcommand{\defeq}{\overset{\mathrm{def}}{=\joinrel=}}
\begin{document}

   \title{Coexisting Tayler instability-driven dynamos in radiative zones: new dynamo solution and its impacts on stellar physics}
   \titlerunning{Coexisting Tayler instability-driven dynamos in radiative zones}

   \author{P. Barrère\inst{1}\fnmsep\thanks{paul.barrere@unige.ch}
          \and
          A. Reboul-Salze\inst{2}
          \and
          P. Eggenberger\inst{1}
          \and
          S. Deheuvels\inst{3,4}
          \and
          C. Rodríguez\inst{1}
          \and
          M. Marchand\inst{1}
          }

         \institute{Département d’Astronomie, Université de Genève, Chemin Pegasi 51, 1290 Versoix, Switzerland
         \and
         Max Planck Institute for Gravitational Physics (Albert Einstein Institute), D-14476 Potsdam, Germany
         \and
         IRAP, Université de Toulouse, CNRS, CNES, UPS, Toulouse, France
         \and 
         Institut Universitaire de France (IUF), Paris, France\\
             }

   \date{Received XXXXX; accepted XXXXX}
 
   \abstract{The recent asteroseismic observations constitute a great challenge for rotating stellar evolution models, which predict too fast internal rotation rates when only hydrodynamic processes are included. This suggests the absence of one or several unidentified angular momentum transport processes in these models. Transport by large-scale and strong magnetic fields in the radiative zone is a promising candidate to explain the observations. While these fields may have a fossil origin, a dynamo driven by the Tayler instability in a shear flow, the so-called Tayler-Spruit dynamo, constitutes a primary mechanism to form the necessary magnetic fields. Despite recent numerical studies, this mechanism remains poorly known. Motivated by this, we investigate the Tayler-Spruit dynamo through a new set of three-dimensional direct numerical simulations. We model the radiative zone as a Boussinesq stably stratified fluid whose differential rotation is maintained by a volumetric body force. We report for the first time the coexistence of two dynamo solutions, which mainly differ by the magnetic field location (near the equator and the polar axis). While the equatorial dynamo is driven by an instability sharing both characteristics of the azimuthal magnetorotational and Tayler instabilities, we mainly investigate the newly identified polar dynamo, which is driven by the standard Tayler instability. We show that this dynamo can still operate and transport angular momentum efficiently in a strong stratification regime, with a Brunt-Väisälä frequency $130$ times larger than the rotation rate. We extract new scaling laws for the different magnetic field components, transport processes, and the minimum shear to trigger the Tayler instability-driven dynamo. Finally, we roughly constrain the signature of the generated magnetic fields on asteroseismic modes propagating in main-sequence and evolved stars. Thus, our results encourage new studies using stellar evolution models including our prescriptions and the search of asteroseismic signals impacted by large-scale azimuthal magnetic fields.}


   \maketitle

\section{Introduction}

The recent asteroseismic data provided by space observatories such as CoRoT~\citep{baglin2006}, TESS~\citep{ricker2015}, and especially Kepler~\citep{borucki2010} allowed for the detection of different oscillation modes (e.g. $g$-modes, $p$-modes, or mixed modes) in thousands of stars, which mostly have low ($\SI{0.5}{M_{\odot}}\lesssim M \lesssim \SI{2}{M_{\odot}}$) or intermediate ($\SI{2}{M_{\odot}}\lesssim M \lesssim \SI{8}{M_{\odot}}$) masses. This led to the release of catalogues composed of (near-)core and, sometimes, surface rotation rates in a wide variety of evolutionary stages~\citep[e.g.][]{mosser2012,deheuvels2014,deheuvels2015,gehan2018,li2020,li2024}. These observational constraints are crucial because rotation significantly impacts the stellar properties and evolution~\citep[e.g.][]{maeder2000,maeder2009}. Besides, the inclusion of rotation and its effects~\citep[e.g. hydrodynamic instabilities, meridional circulation][]{zahn1992} in 1D evolution models provide more realistic grids of stellar evolution~\citep[e.g.][]{ekstrom2012}. However, asteroseismic constraints unambiguously reveal that stellar core rotations are still slower by several orders of magnitude (e.g. three for red giants) than predicted by rotating 1D models~\citep[e.g.][]{eggenberger2012,marques2013,ceillier2013,ouazzani2019}. Therefore, additional physical processes extracting angular momentum (AM) efficiently from stellar cores must be included in the models to fit the observations. Furthermore, the needed AM transport efficiency has been quantified for different evolution stages in low- and intermediate-mass stars and white dwarfs~\citep[e.g.][]{eggenberger2017,eggenberger2019a,denhartogh2019,moyano2022}.

The proposed new AM transport mechanisms rely on two missing ingredients in rotating 1D stellar evolution models: either internal waves, or magnetic fields. First, internal gravity waves triggered by convective plumes at the interface between radiative and convective regions can deposit AM in the damping regions. The trigger and the efficiency of this process were thoroughly investigated analytically and numerically~\citep[e.g.][]{rogers2013,fuller2014,pincon2016}. Moreover, \citet{belkacem2015b,belkacem2015a} and \citet{bordadagua2025} proposed mixed oscillation modes as a promising candidate to explain the needed AM transport in the upper part of the red giant branch. However, on the one hand, internal gravity wave-driven transport is inefficient in red giants~\citep{pincon2017}, and on the other hand, mixed-modes cannot explain the rotation of sub-giants and early red giants. 

Second, large-scale magnetic fields can transport AM via Maxwell stresses. In stellar interiors, two distinct magnetic field formation scenarios are expected. On the one hand, they can be fossil fields, that is, amplified by magnetic flux conservation during the collapse of the initial molecular cloud or generated by dynamo action in the early convective core~\citep[e.g.][]{takahashi2021,skoutnev2025}. On the other hand, they can be amplified and sustained by one or several acting dynamo mechanisms. The presence of magnetic fields in radiative zones, even though expected, is now confirmed by recent asteroseismic studies of red giants~\citep{li2022,li2023,deheuvels2023,hatt2024}. These observations provide important constraints on the magnetic field intensity and geometry. Indeed, the fields are in the order of $10^4-10^5\SI{}{G}$ with a dominant radial component, which is not necessarily consistent with a pure magnetic dipole. The detected field strengths are in global agreement with a fossil field stemming from a convective core dynamo in the early evolutionary stages, as seen in numerical simulations~\citep{brun2005,augustson2016,augustson2019}. However, we could expect too strong magnetic fields to couple the core and the envelope, suppressing differential rotation, despite core contraction during late stages. This is in tension with asteroseismic analysis because the observed magnetised red giants show common rotational properties, and so differential rotation. To temper this argument, note that the efficiency of transport by fossil fields, also called magnetic webs, derived by~\citet{skoutnev2025}, shows that the core rotation of red giants can be matched, but for relatively low overshoot parameters in evolution models. Another remaining uncertainty is the magnetic field geometry after its relaxation to a stable configuration. Despite many analytical and numerical studies bringing a better understanding of the stability conditions~\citep{braithwaite2008,duez2010a,duez2010b,becerra2022a,becerra2022b}, many important ingredients are still lacking in these models, such as the star rotation and an initial magnetic field configuration stemming from a saturated dynamo state.

The detected magnetic fields in red giants are expected to be localised where the observations of mixed modes are more sensitive. While the sensitivity is maximum inside the hydrogen-burning shell (HBS), the detections are also sensitive to the layers beneath it~\citep{li2022}. Therefore, the presence of weaker radial magnetic fields in the rest of the radiative zone is not excluded. Furthermore, magnetic fields may exist in stars where the current detection methods remain insensitive. If not formed in the early evolutionary phases, these fields could be generated by dynamo action in the radiative regions. These zones are stably stratified, that is, the temperature and the chemical gradients prevent the development of convective motions. The dynamo must, therefore, be driven by MHD instability-generated turbulence. The mechanism can also combine these instabilities with differential rotation, which shears the poloidal magnetic field into a toroidal geometry. Several MHD instabilities driving dynamo action have been studied: the magnetic buoyancy~\citep{cline2003}, the magnetorotational instability~\citep[MRI, e.g.][]{reboul2021a,reboul2022,guilet2022}, and the Tayler instability~\citep[e.g.][]{spruit2002,denissenkov2007,zahn2007,fuller2019}. The first one stems from the tendency of magnetised fluid to be `lighter' than its non-magnetised surrounding~\citep{parker1955a} and was invoked as a possible contributor to the solar dynamo~\citep{vasil2008,vasila2009,duguid2023}. 
The MRI is an MHD instability that feeds off differential rotation~\citep{balbus1991,hawley1996}. The few numerical studies of this instability for stellar radiative zones show that the development of MRI is favoured by latitudinal differential rotation, which can appear even for relatively strong stratifications~\citep{jouve2020,gouhier2021,gouhier2022}. \citet{meduri2024} provide a scaling law for the diffusion coefficient associated with the transport, which is calibrated on simulations of transient MRI-driven turbulence as a function of rotation and stratification. This law has not been implemented in 1D stellar models yet, and current evolution models only rely on simplistic formulas~\citep[e.g.][]{wheeler2015,spada2016,griffiths2022,moyano2023}.

\begin{figure*}
    \centering
    \includegraphics[width=0.75\linewidth]{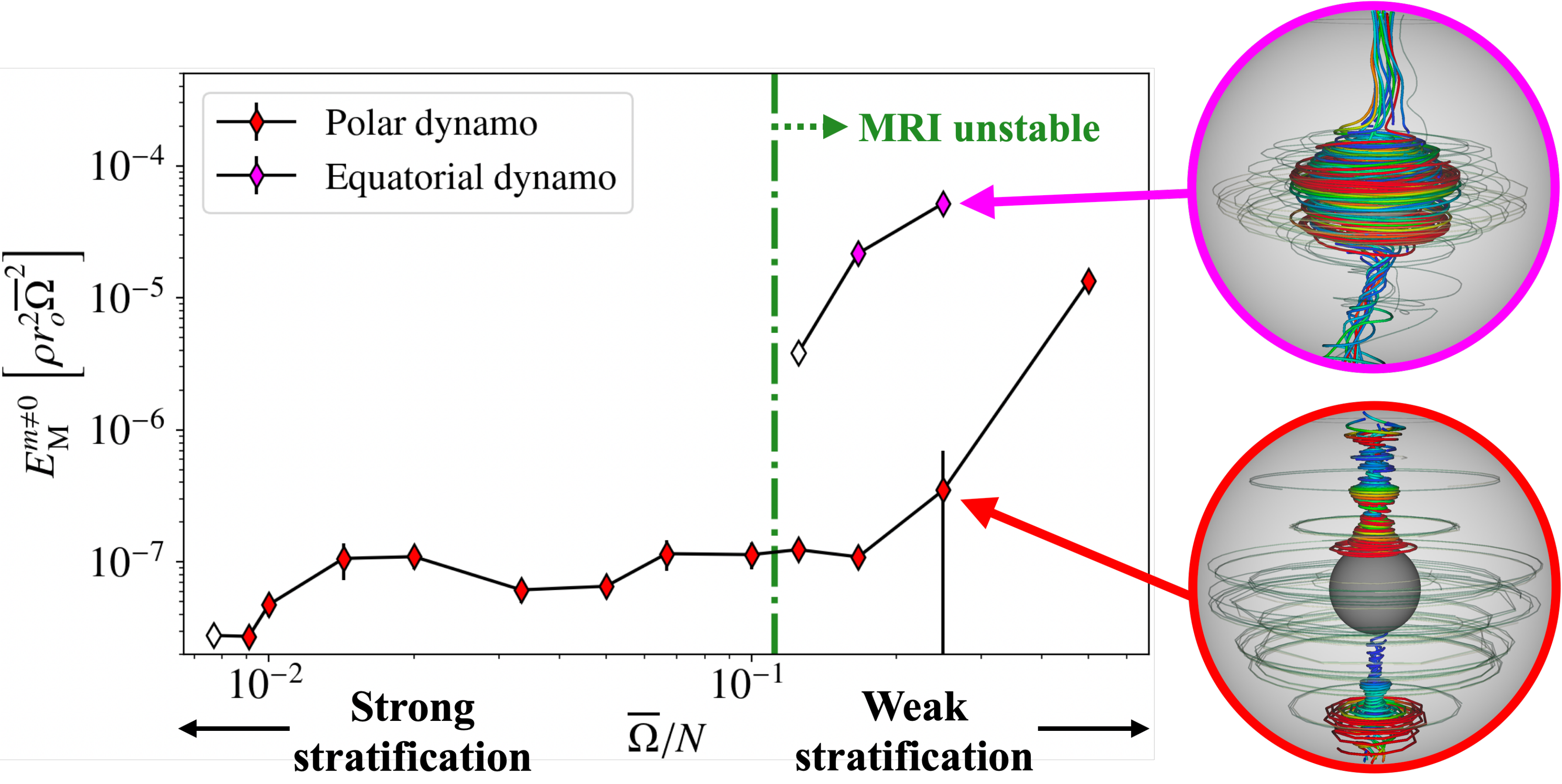}
    \caption{Bifurcation diagram of the time- and volume-averaged turbulent magnetic energy as a function of the ratio of the frame rotation rate to the Brunt-Väisälä frequency. The error bars indicate the standard deviation. The red and magenta markers represent two distinct Tayler-Spruit dynamos characterised by Tayler modes near the polar axis (polar dynamo) and at the equator (equatorial dynamo), respectively. The empty markers indicate transient behaviour. 3D representations of the magnetic field lines for both dynamos at $\overline{\Omega}/N=0.25$ are plotted on the right. The colour indicates the magnetic strength, which is around $\SI{e5}{G}$ for the hydrogen-burning shell of red giants. Additional grey magnetic field lines are also plotted to display the dominant toroidal magnetic field produced by the dynamo.}
    \label{fig:bifurcation}
\end{figure*}

In this work, we will focus on the Tayler instability-driven dynamo, also called the Tayler-Spruit dynamo. The Tayler instability is a purely magnetic instability that feeds off strong toroidal magnetic fields~\citep{tayler1973,goossens1981}. \citet{spruit2002} proposed that the Tayler modes could be sheared to regenerate the initial toroidal magnetic field and close a dynamo loop. To correct some inconsistencies~\citep[see][]{denissenkov2007,zahn2007}, \citet{fuller2019} revised the first model of the Tayler-Spruit dynamo and concluded that the mechanism is very efficient at transporting AM, even for the strong stratifications of the red giant HBS. Despite the absence of numerical evidence to confirm the existence of the dynamo, both transport prescriptions derived by~\citet{spruit2002} and~\citet{fuller2019} have been implemented in several stellar evolution codes~\citep[e.g.][]{maeder2003,heger2005,eggenberger2005,eggenberger2019c}. The magnetised evolution models gave several indications about the impact of the dynamo-induced transport on stellar evolution. Both analytical prescriptions can reproduce the solar chemical abundances and most of the radiative zone rotation, but robust data for the solar core rotation are still lacking to differentiate between the dynamo models~\citep{eggenberger2022}. However, the original model is not efficient enough to reproduce the late stages of low- and intermediate-mass stars~\citep[e.g.][]{cantiello2014,denhartogh2019,eggenberger2022b}. The formalism proposed by~\citet{fuller2019} reproduces well the rotation of red giants and helium-burning stars. A version of the Tayler-Spruit dynamo calibrated to match red giant rotation~\citep[derived by][]{eggenberger2022b} also reproduces the rotation of main-sequence $\gamma$-Dor stars~\citep{moyano2023}. Nonetheless, this formalism has trouble matching the rotation of subgiants~\citep{eggenberger2019b}. Moreover, the AM transport becomes too efficient to explain the ratio of the convective core to near-core rotation rates in $\gamma$-Dor stars~\citep{moyano2024}, and to reproduce the rotation of white dwarfs~\citep{denhartog2020}.

\citet{petitdemange2023} provided the first identification of the Tayler-Spruit dynamo in 3D direct numerical simulations of stellar radiative zones. The measured magnetic AM transport is consistent with the scaling law of the original Tayler-Spruit dynamo model~\citep{petitdemange2024}. To alleviate the tension between this dynamo model and the observations of post-main-sequence stars, \citet{daniel2023} argued that the additional transport required can be done by Reynolds stresses. While they forced differential rotation by imposing fixed rotation rates at the spherical boundaries (spherical Taylor-Couette configuration), we will use a volumetric forcing. This configuration is less prone to hydrodynamic instabilities, which facilitates the identification of the acting dynamo in the simulation. In this paper, we demonstrate the existence of a new Tayler-Spruit dynamo in a setup relevant for stellar stably stratified zones, and characterise its impact on stellar physics. This new solution recalls one of the Tayler-Spruit dynamo branches identified in proto-neutron stars by~\cite{barrere2023,barrere2025,barrere2026a}, and can be maintained for the extreme stratifications observed in evolved stars. 

In the following Sect.~\ref{sec:methods}, we describe the numerical setup and methods. Sects.~\ref{sec:branch} and~\ref{sec:laws} present the new dynamo solution in different regimes of stratification and the extracted scaling laws of the magnetic field and the transport processes, while we show the different implications for observations in Sect.~\ref{sec:applications}. Finally, we discuss the limits of the methods and results in Sect.~\ref{sec:discussion}, and draw the conclusions in Sect.~\ref{sec:conclusions}.

\section{Methods}\label{sec:methods}

We model a stellar radiative region as a stably stratified and Boussinesq MHD flow evolving between two concentric spheres of radius $r_i$ and $r_o = 4r_i$, defining the sphere gap $d= r_o-r_i = 0.75r_o$. We apply no-slip and electrically insulating conditions on both shells. For every simulation, we assume fixed and uniform kinematic viscosity $\nu$, thermal diffusivity $\kappa$, and magnetic diffusivity $\eta$, which are characterised by the thermal and magnetic Prandtl numbers:
\begin{align}
    &Pr\defeq \frac{\nu}{\kappa}=0.1\,,\\
    &Pm \defeq \frac{\nu}{\eta}=4\,,
\end{align}
respectively. In line with the Boussinesq approximation, the fluid density $\rho$ is uniform, which implies a gravity proportional to the radius: $\mathbf{g}=-g_or/r_o\mathbf{e}_r$, where $g_o$ is the gravitational acceleration at the outer sphere $g_o$. The stable stratification is imposed by fixing $\Delta T = T_o-T_i >0$ and is represented by the Rayleigh number:
\begin{equation}
    Ra \defeq \frac{d^4N^2}{\nu\kappa} = \frac{d^3\alpha g_o\Delta T}{\nu\kappa}\in\left[\SI{4e9}{},\SI{4e11}{}\right]\,,
\end{equation}
where $N$ and $\alpha$ are the Brunt-Väisälä frequency and thermal expansion coefficient, respectively. The rotation is characterised by the Ekman number:
\begin{equation}
    E \defeq \frac{\nu}{d^2\overline{\Omega}} \in\left[\SI{1e-5}{},\SI{6.5e-5}{}\right]\,,
\end{equation}
with $\overline{\Omega}$ the rotation rate of the frame, which corresponds to the rotation rate at $r\approx0.8r_o$ in the latitudinally-averaged rotation profiles. Finally, instead of the Rayleigh number $Ra$, we will use the ratio of the frame rotation rate to the Brunt-Väisälä frequency to characterise the stratification
\begin{equation}
    \overline{\Omega}/N = \sqrt{\frac{Pr}{Ra E^2}} \in [0.0077,0.5]\,.
\end{equation}

\subsection{Governing equations}
These numbers are found in the Boussinesq MHD equations by scaling the length in units of the sphere gap $d$, the time in units
of viscous time $d^2/\nu$, the magnetic field in units of $(4\pi\rho\eta\overline{\Omega})^{1/2}$, and the temperature in units of the temperature contrast between the two spheres $\Delta T$. These equations describe the coupled evolution of the velocity $\mathbf{v}$ and magnetic field $\mathbf{B}$, and read:
\begin{align} \label{eq:1}
    D_t\mathbf{v} & = -\nabla p' -\frac{2}{E}\mathbf{e}_z\times\mathbf{v} - \frac{Ra}{Pr}T' \mathbf{e}_r \nonumber\\
    & + \frac{1}{E\,Pm} (\nabla\times \mathbf{B})\times \mathbf{B} + \Delta\mathbf{v} + \mathbf{f}\,, \\ \label{eq:3}
    D_t T' &+ \mathbf{v}\cdot\nabla \overline{T} =\frac{1}{Pr}\Delta T'\,,\\ \label{eq:4}
    \partial_t\mathbf{B} &=\nabla\times(\mathbf{v}\times\mathbf{B})+\frac{1}{Pm}\Delta \mathbf{B}\,,\\ \label{eq:5}
    \nabla \cdot \mathbf{v} &=0\,,\:\nabla\cdot\mathbf{B} =0\,, 
\end{align}
where $p'$ is the reduced pressure (i.e. the pressure divided by the density), and the temperature field is the addition of the temperature of the reference state $\overline{T}(r)$ and its fluctuation $T'(r,\theta,t)$. $\mathbf{e}_z$ and $\mathbf{e}_r$ are the unit vectors of the axial and the spherical radial directions, respectively. Finally, $\mathbf{f}$ is an additional body force for the volumetric forcing (see Sect.~\ref{ssec:volForce}). Note that the presence of the chemical composition and local heat sources is ignored in these equations.

\subsection{Volumetric forcing}\label{ssec:volForce}
To force the differential rotation in our simulations, we add an axisymmetric forced contribution $v_{\rm f}=r\sin\theta\Omega_{\rm f}$ to the azimuthal velocity field, such that the velocity field $\mathbf{v}=\mathbf{u}+v_{\rm f}\mathbf{e}_{\phi}$. $v_{\rm f}$ is the stationary field ($D_t(v_{\rm f}\mathbf{e}_{\phi})=0$) towards which the axisymmetric azimuthal velocity field $u_{\phi}^{m=0}$ relaxes within a timescale $\tau^{-1}$. Hence, the additional dissipation term
\begin{equation}
    \mathbf{f}\defeq-\tau u_{\phi}^{m=0}\mathbf{e}_{\phi}\,,
\end{equation}
in the momentum equation (Eq.~\ref{eq:3}). In viscous units, the relaxation time is fixed at $\tau^{-1}=10^{-4}\approx2-10\times E$. Note that for the strongly stratified case at $\overline{\Omega}/N=\SI{1.4e-2}{}$, the dynamo can be maintained for $\tau^{-1}\lesssim 29\times E$ (see Appendix~\ref{sec:forcingTime}).

Since we explore strongly stratified regimes, we choose a shellular forced rotation, that is only dependent on the radius:
\begin{equation}\label{eq:rotation}
    \Omega_{\rm f}(r) = \frac{\Omega_i}{\left(1+\left(r/r_i\right)^{20 q_o}\right)^{1/20}}\,,
\end{equation}
where $q_o$ corresponds to the shear rate $q=(r/\Omega)d_r\Omega$ of the forced contribution and is fixed to $1$. $\Omega_i$ is the rotation rate at the inner boundary and is chosen so that the ratio of total AM over the moment of inertia is equal to the frame rotation rate $\overline{\Omega}$. Numerical simulations without magnetic fields show that the flow remains hydrodynamically stable with the chosen forced profile. The stationary axisymmetric velocity field obtained in these non-magnetic simulations is then used to initialise the MHD ones.

\subsection{Numerical methods}
To integrate Eqs.~\eqref{eq:1}--\eqref{eq:5} in 3D spherical geometry, we use the open-source pseudo-spectral code MagIC (commit 2266201a5) \citep{wicht2002,gastine2012,schaeffer2013}. The velocity and magnetic fields are decomposed into poloidal and toroidal components:
\begin{align}
    \mathbf{v} &= \nabla \times \nabla \times (W\mathbf{e}_r) + \nabla\times(Z\mathbf{e}_r)\,,\\
    \mathbf{B} &= \nabla \times \nabla \times (b\mathbf{e}_r) + \nabla\times(a_j\mathbf{e}_r)\,,
\end{align}
where $W$ and $Z$ are the respective poloidal and toroidal kinetic potentials, while $b$ and $a_j$ are the magnetic ones. The horizontal (i.e. in colatitude $\theta$ and longitude $\phi$) and radial dependencies of these fields and reduced pressure $p'$ are then expanded into spherical harmonics and Chebyshev polynomials. For the time stepping, we use an implicit-explicit Runge-Kutta scheme developed by~\citet{boscarino2013}. The resolution is varied between the simulations and can be found in Table~\ref{tab:sim_input}. The numerical simulations with the lowest $\overline{\Omega}/N\in\{0.25,0.5\}$ are initiated with either a purely $(\ell=1,m=0)$ or $(\ell=2,m=0)$ poloidal magnetic field that are potential fields at the core ($b\propto 1/r$) and the outer boundary ($b\propto (r/r_i)^3(1-5/7\times(r/r_o)^2)$), respectively. Random noise is added to the velocity field at the start of the simulation. The rest of the simulations are initialised by the nearby saturated state of a run with a weaker stratification. Using this procedure, $\overline{\Omega}/N$ is increased gradually to avoid losing the dynamo. 

\subsection{Outputs}\label{sec:output}
In Sect.~\ref{sec:branch}, except for in Fig.~\ref{fig:lTI}, the outputs are rescaled to be in rotational units using $r_o$ and $P\equiv 2\pi/\overline{\Omega}$ as length and time units. The energies are volume- and time-averaged in the time interval of the saturated dynamo state for the bifurcation diagram (Fig.~\ref{fig:bifurcation}). 

For Fig.~\ref{fig:lTI} in Sect.~\ref{sec:branch} and the scaling laws in Sect.~\ref{sec:laws}, the radial length scale of the Tayler modes ($l_{\rm TI}$), the magnetic field strengths ($B_{\phi}^{m=0}$, $B_{r}^{m=0}$, $B_{\rm tot}^{m\neq0}$, $B_{\perp}^{m\neq0}$, $B_{r}^{m\neq0}$) and viscosities associated to the different transport mechanisms ($\nu_{\rm M}$, $\nu_{\rm R}$, $\nu_{\rm mix}$) are also scaled in rotational units but using the following local quantities: the shear rate ($q$), local radius ($r_{\rm loc}$) and rotation rate ($\Omega_{\rm loc}$). Note that the dimensionless magnetic fields are the equivalent of the Lehnert number, which characterises the ratio of the Lorentz to the Coriolis force:
\begin{equation}
    2Le\defeq \frac{B}{\sqrt{4\pi\rho r_{\rm loc}^2\Omega_{\rm loc}^2}}\,.
\end{equation}
All these quantities are measured locally as described in Appendix~\ref{sec:average} (Figs.~\ref{fig:lTI}--\ref{fig:AMT}). Finally, the values of every quantity we introduce and use in the following plots of this paper are listed in Tables.~\ref{tab:sim_input}-~\ref{tab:sim_nu}.

\section{New solution of the Tayler-Spruit dynamo}\label{sec:branch}

\subsection{Two co-existing dynamos}\label{sec:bistable}

\begin{figure}
    \centering
    \includegraphics[width=\linewidth]{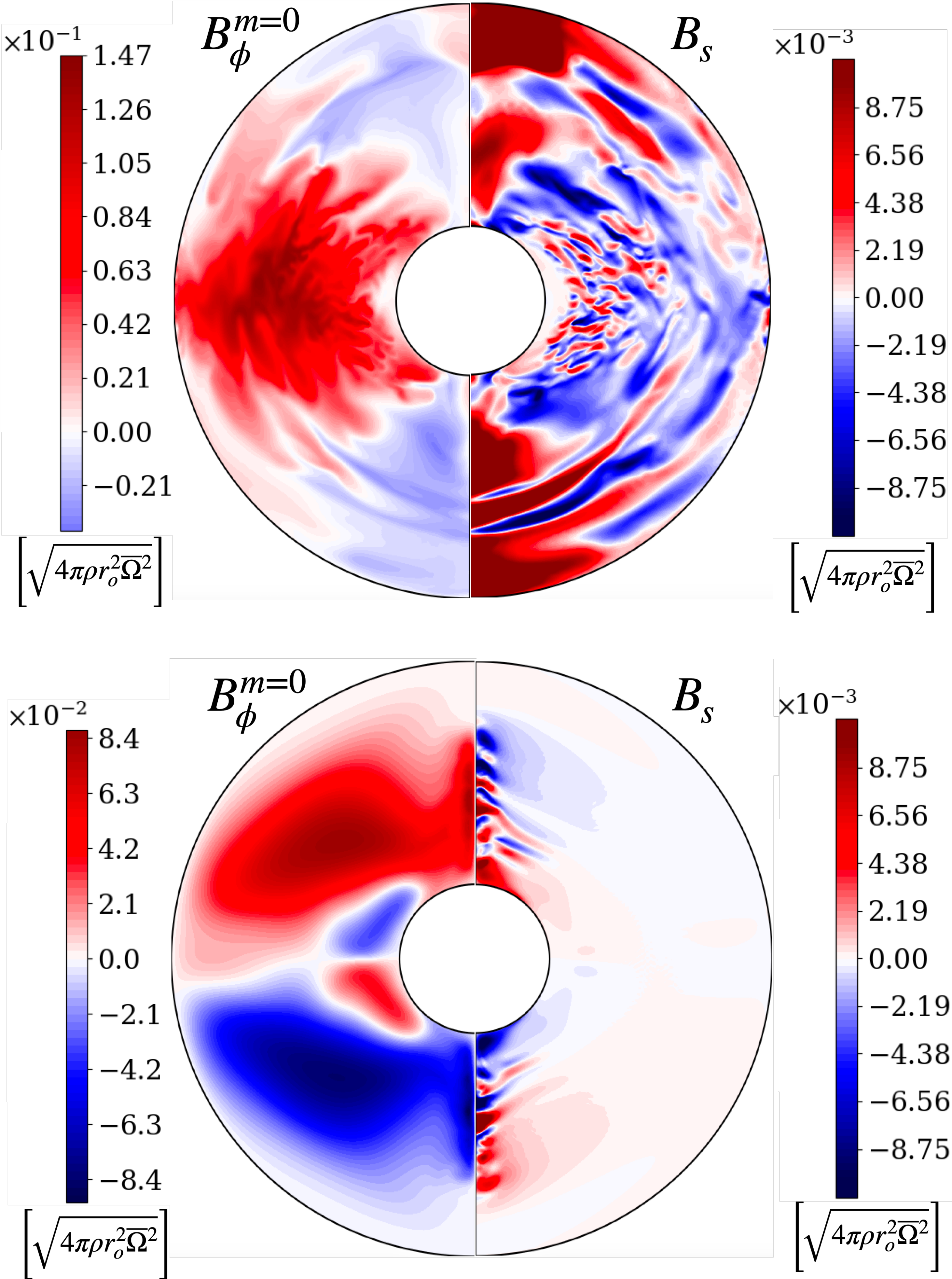}
    \caption{Meridional slices of the axisymmetric azimuthal and the $s=r \sin{\theta}$--component of the magnetic fields (left and right, respectively) for the equatorial (top) and the polar (bottom) dynamos at  $\overline{\Omega}/N=0.25$.}
    \label{fig:slices}
\end{figure}

\begin{figure*}
    \centering
    \includegraphics[width=0.85\linewidth]{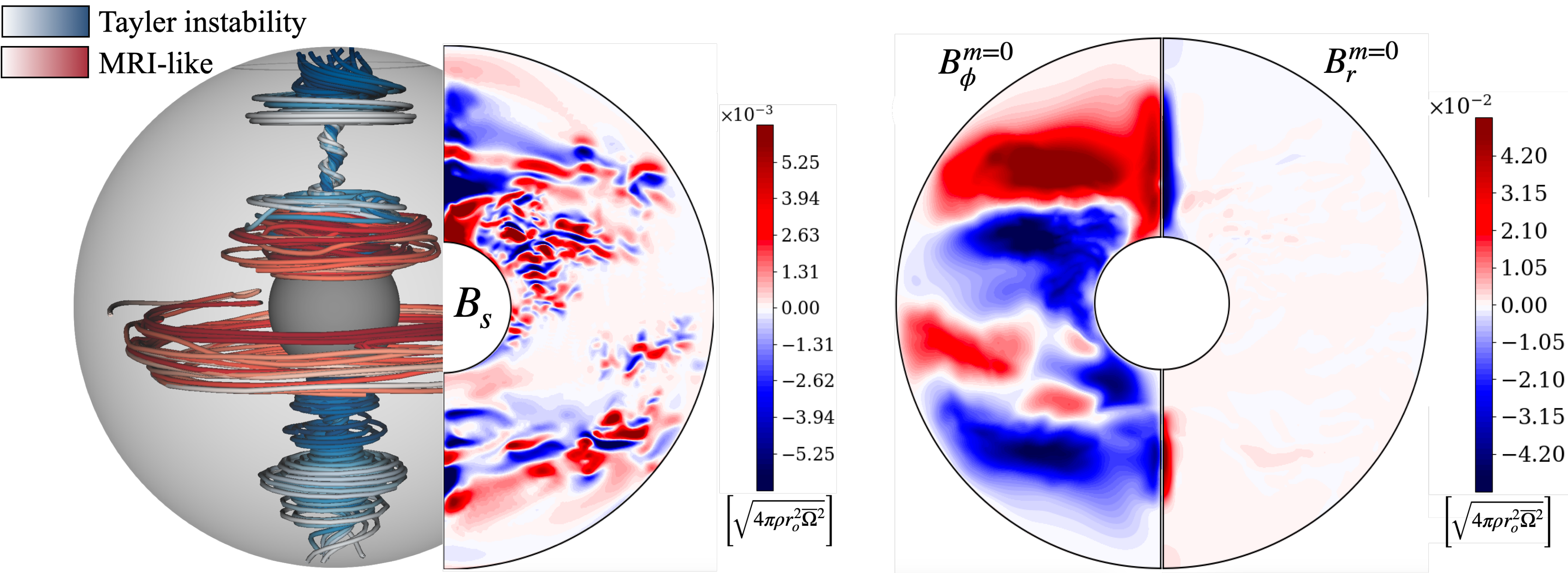}
    \caption{Left: 3D snapshots of the magnetic field lines, coloured depending on the instability they undergo (Tayler in blue, MRI-like in red). A meridional slice of the $s=r \sin{\theta}$--component of the magnetic field is also plotted on the right. Right: Meridional slices of the axisymmetric azimuthal and radial magnetic fields. These snapshots are extracted from the simulation of the Tayler-Spruit dynamo at $\overline{\Omega}/N=0.5$.}
    \label{fig:TaylerMRI}
\end{figure*}

The different obtained dynamo states are gathered in the bifurcation diagram displayed in Fig.~\ref{fig:bifurcation}, which represents the non-axisymmetric magnetic energy of the solutions as a function of the input ratio $\overline{\Omega}/N$. Two distinct dynamos co-exist and can be reached depending on the initial magnetic field geometry:
\begin{itemize}
    \item[(i)] For an initial ($\ell=2,m=0$) poloidal field, we obtain the solutions coloured in magenta, which can be observed for $\overline{\Omega}/N\in[0.13,0.25]$. As displayed in the 3D snapshot on top in Fig.~\ref{fig:bifurcation} and the meridional slice of $B_{\phi}^{m=0}$ on top in Fig.~\ref{fig:slices}, the toroidal field generated by the dynamo (shear and electromotive force) is focused on the equatorial plane, hence the name `equatorial dynamo'. The meridional slice of $B_s$ on top in Fig.~\ref{fig:slices} shows that unstable modes develop near the inner spherical boundary, where the radial gradient of $B_{\phi}^{m=0}$ is globally positive. \citet[][]{petitdemange2023,petitdemange2024} observed a similar dynamo in a simulated radiative zone. Note that the magnetic field that can be seen outside the vicinity of the equatorial plane is the remnant of an initial transient, where unstable modes also developed near the polar axis.
    
    \item[(ii)] For an initial ($\ell=1,m=0$) poloidal field, we obtain the dynamo coloured in red, which can be observed for the much larger interval $\overline{\Omega}/N\in[0.0077,0.5]$. In this case, the toroidal field remains strong in most of the integrated volume but with the opposite equatorial symmetry compared to the equatorial dynamo, as $B_{\phi}^{m=0}$ tends to $0$ towards the equator. Also, the unstable modes are located around the polar axis, and not the equatorial plane, hence the name `polar dynamo'.
\end{itemize}

Both solutions are obtained for a flow that is stable to convection and to hydrodynamic instabilities, indicating the action of two MHD instability-driven subcritical dynamos. In both cases, the most unstable non-axisymmetric mode is $m=1$ (see the spectra in Appendix~\ref{sec:spectra}), and the magnetic field lines are mostly toroidal (see the 3D snapshots in Fig.~\ref{fig:bifurcation}). The MHD instabilities sustaining both dynamos are driven by the magnetic pressure. However, different components of this pressure dominate depending on the dynamo: the radial and latitudinal components drive the equatorial dynamo, while the latitudinal and longitudinal ones drive the polar dynamo. This suggests that the nature of the instability may be different between both dynamos. The instabilities that can develop in a stably-stratified flow with a dominant azimuthal magnetic field are the azimuthal MRI ~\citep[][]{ogilvie1996,rudiger2007} and the Tayler instability~\citep[][]{tayler1973,goossens1981}, which feed off differential rotation and strong vertical currents, respectively. As both differential rotation and strong vertical currents are present in our simulations, the identification of the driving instability for both dynamos is not straightforward.

Based on the analysis in Appendix~\ref{sec:tests}, which studies the linear evolution of ($\ell=1,m=0$)-toroidal magnetic field instabilities, we argue that the polar dynamo is driven by the Tayler instability. Indeed, in the absence of differential rotation, a dominating $m=1$ mode grows around the rotation axis and so feeds off vertical currents only. The correlation between the unstable mode location and the regions with positive latitudinal gradients of the toroidal fields also supports this assertion~\citep{goossens1980a}. Note that the location and the geometry of the magnetic field recall the strong Tayler-Spruit dynamo reported by~\citet{barrere2023,barrere2025} for proto-neutron stars spun-up by fallback, where the shear rate is positive ($q>0$).

On the other hand, according to simulations of a differentially-rotating flow in Appendix~\ref{sec:tests}, unstable $m=1$ or $m=2$ modes develop at near the equator. Their location suggests that they drive the equatorial dynamo. Since it appears only in differentially rotating (but only $q<0$) and non-zero current regions, we interpret the instability as a mix of both azimuthal MRI and standard Tayler instability. This type of instability was analytically predicted by~\citet{kirillov2014} for a cylindrical Taylor-Couette flow. Interestingly, we find that its growth rate depends on rotation and initial magnetic field strength in a similar way as the polar Tayler instability (see Fig.~\ref{fig:growth}). Besides, note that the equatorial dynamo respects the stability criterion of MRI~\citep{balbus1991,balbus1998,menou2004}
\begin{equation}\label{eq:MRIcrit}
    -q < \frac{N_{\rm eff}^2}{2\overline{\Omega}^2} = \frac{\eta}{\kappa}\frac{N^2}{2\overline{\Omega}^2}\,,
\end{equation}
which predicts that the flow is unstable to MRI for $\overline{\Omega}/N\gtrsim 0.11$ (green vertical line in Fig.\ref{fig:bifurcation}), for the initial shear rate $q=-1$. In this equation, we introduce the effective Brunt-Väisälä frequency $N_{\rm eff}$ that characterises stratification including diffusive effects. Note that stratification restricts movements along the differential rotation, which are necessary for the equatorial MHD instability, while differential rotation is not necessary for polar Tayler instability.

\subsection{Impact of stable stratification}

The Tayler-Spruit dynamo is present in our simulations over almost two orders of magnitude of $\overline{\Omega}/N$, which allows us to investigate the impact of stratification on the magnetic field geometry. For a weak stratification ($\overline{\Omega}/N=0.5$), the 3D magnetic lines and the meridional slice of $B_s$ in Fig.~\ref{fig:TaylerMRI} show the presence of small-scale magnetic field around the equatorial plane, in addition to the Tayler modes on the polar axis. The instability producing these equatorial modes shares similarities with the azimuthal MRI. Indeed, while $B_{\phi}^{m=0}$ is initially with a geometry ($\ell=2,m=0$), we observe local reversals around the equatorial plane, creating a more complex geometry (see meridional slice of $B_{\phi}^{m=0}$ in Fig.~\ref{fig:TaylerMRI}). Moreover, $B_r^{m=0}$ is much weaker near the equator than close to the polar axis, where the Tayler-Spruit dynamo acts (see meridional slice of $B_{r}^{m=0}$ in Fig.~\ref{fig:TaylerMRI}). Finally, the instability is driven by the radial and latitudinal magnetic pressure, and appears only when the stability criterion (Eq.~\ref{eq:MRIcrit}) is respected. These indications are characteristic of the MRI that has already been studied in a spherical configuration~\citep{reboul2021a,reboul2022,meduri2024}. We must note that it is unclear whether this instability also taps into the energy associated with the toroidal magnetic field, like the instability driving the equatorial dynamo, because we only observe this instability in its nonlinear regime.

\begin{figure}
    \centering
    \includegraphics[width=\linewidth]{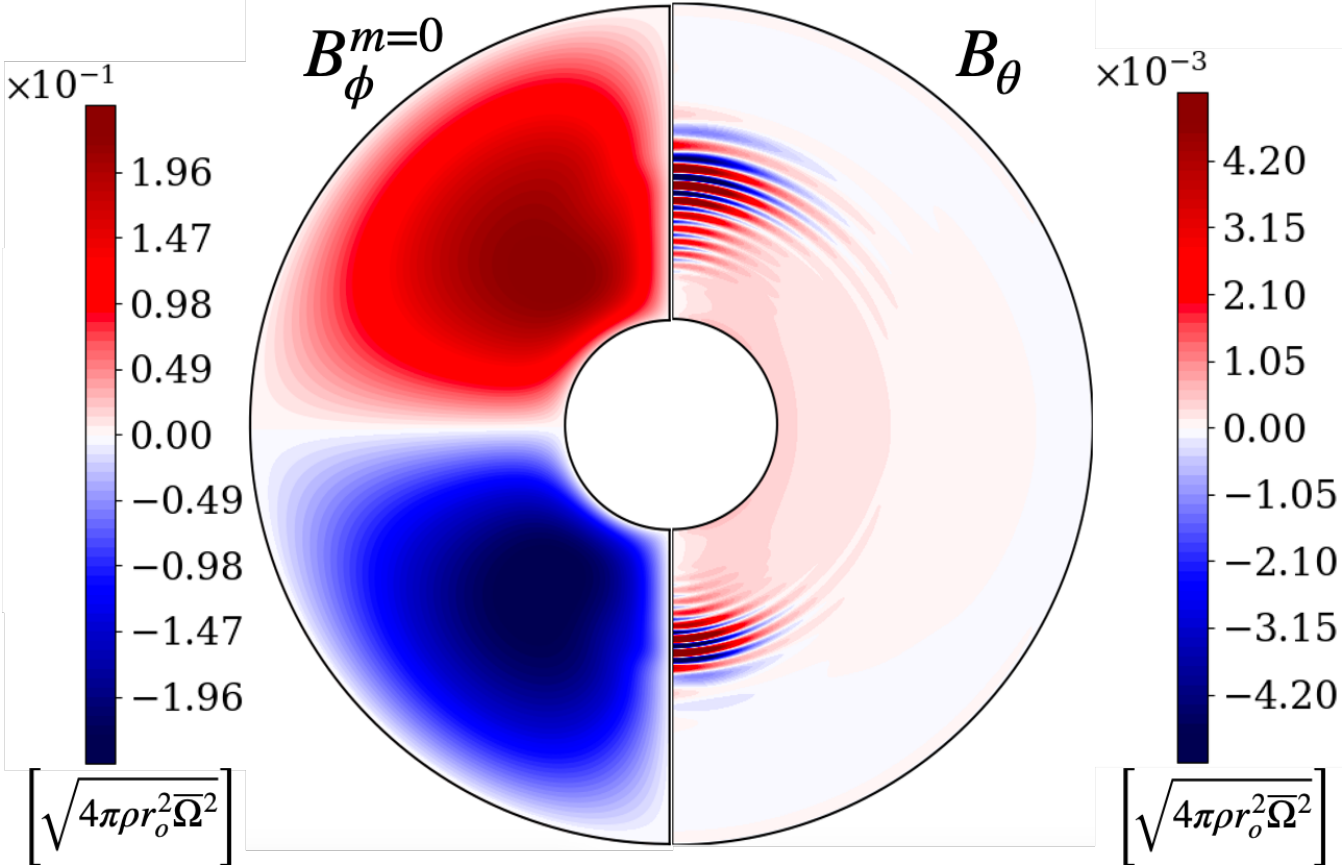}
    \caption{Meridional slices of the axisymmetric azimuthal and total latitudinal magnetic fields (left and right, respectively) of the Tayler-Spruit dynamo at $\overline{\Omega}/N=0.0077$.}
    \label{fig:NO130}
\end{figure}

As seen in Fig.~\ref{fig:NO130}, a consequence of the absence of MRI is the geometry of $B_{\phi}^{m=0}$ that becomes ($\ell=1,m=0$), and its dynamics is stationary, like in previous studies of the Tayler-Spruit dynamo with $q>0$~\citep{barrere2023,barrere2025,barrere2026a}. While the geometry of $B_{\phi}^{m=0}$ remains the same for a wide range of $\overline{\Omega}/N\in[0.0077,0.25]$, the most striking impact of stable stratification is the reduction of the radial length scale of Tayler modes. This is clearly shown when we compare the meridional slices of $B_s$ at $\overline{\Omega}/N=0.25$ in Fig~\ref{fig:TaylerMRI} and of $B_{\theta}$ at $\overline{\Omega}/N=0.0077$ in Fig.~\ref{fig:NO130}. We quantify the radial length scale of the Tayler modes $l_{\rm TI}$ in Fig.~\ref{fig:lTI} (black stars) by using time-averaged profiles of the non-axisymmetric radial magnetic field close to the polar axis ($\theta\approx 2^{\circ}$). In the same figure, we also plotted the upper and lower limits constraining $l_{\rm TI}$ derived by~\citet{spruit1999}:
\begin{equation}\label{eq:bound}
    \eta \frac{\Omega_{\rm loc}}{\omega_{\rm A}^2} \lesssim  \frac{l_{\rm TI}}{r_{\rm loc}} \lesssim \frac{\omega_{\rm A}}{N_{\rm eff}}\,,
\end{equation}
where we assumed that the horizontal length scale of the modes is $l_{\perp}\sim r_{\rm loc}\sim r_o/2$. We also introduce the local quantities $r_{\rm loc}$, $\Omega_{\rm loc}$ (see Sect.~\ref{sec:output}), and the Alfvén frequency 
\begin{equation}
    \omega_{\rm A}\defeq\frac{B_{\phi}^{m=0}}{\sqrt{4\pi\rho r_{\rm loc}^2}}\,.
\end{equation}
The calculation of the last three quantities is described in Appendix~\ref{sec:average}. Fig.~\ref{fig:lTI} shows that the measured $l_{\rm TI}$ is well constrained by the theoretical constraints in Eq.~\ref{eq:bound}, but multiplied by a factor $4$, which is reasonable since the limits in Eq.~\ref{eq:bound} are orders of magnitude. As predicted theoretically, we lose the dynamo state when the values of both limits are close. Our simulations therefore support the prediction of the critical toroidal magnetic field above which it becomes Tayler unstable
\begin{equation}\label{eq:critic}
    \omega_{\rm A,c} = \Omega_{\rm loc}\left(\frac{N_{\rm eff}}{\Omega_{\rm loc}}\right)^{1/2}\left(\frac{\eta}{r_{\rm loc}^2\Omega_{\rm loc}}\right)^{1/4}\,,
\end{equation}
which is obtained by equating both limits in Eq.\ref{eq:bound}~\citep{spruit1999}.

\section{Scaling laws}\label{sec:laws}
To predict the impact of the Tayler-Spruit dynamo on stellar evolution, we must determine the scaling laws followed by the saturated magnetic field (Sect.~\ref{sec:mag_laws}) and the different transport mechanisms (Sect.~\ref{sec:amt_laws}). To this end, we measure these different quantities by first calculating their time and horizontally-averaged radial profiles. Then, in this profile, we average the quantity between two radii $r_{\rm min}$ and $r_{\rm max}$, constraining the region where the dynamo generates most of the magnetic energy, as explained in Appendix~\ref{sec:average}. This method of measurement is relevant in the case of scaling laws for 1D stellar evolution models, in which the quantities do not depend on the horizontal directions. From the scaling laws of the magnetic field, we then determine a new prescription for the minimum shear rate that is necessary for the Tayler instability to occur. We also confront the calculated scaling laws with previous analytical investigations of~\citet{spruit2002} and~\citet{fuller2019}, which are in global agreement with the direct numerical simulations of~\citet{petitdemange2024} and~\citet{barrere2025}, respectively.

\subsection{Magnetic field}\label{sec:mag_laws}
\begin{figure}
    \centering
    \includegraphics[width=\linewidth]{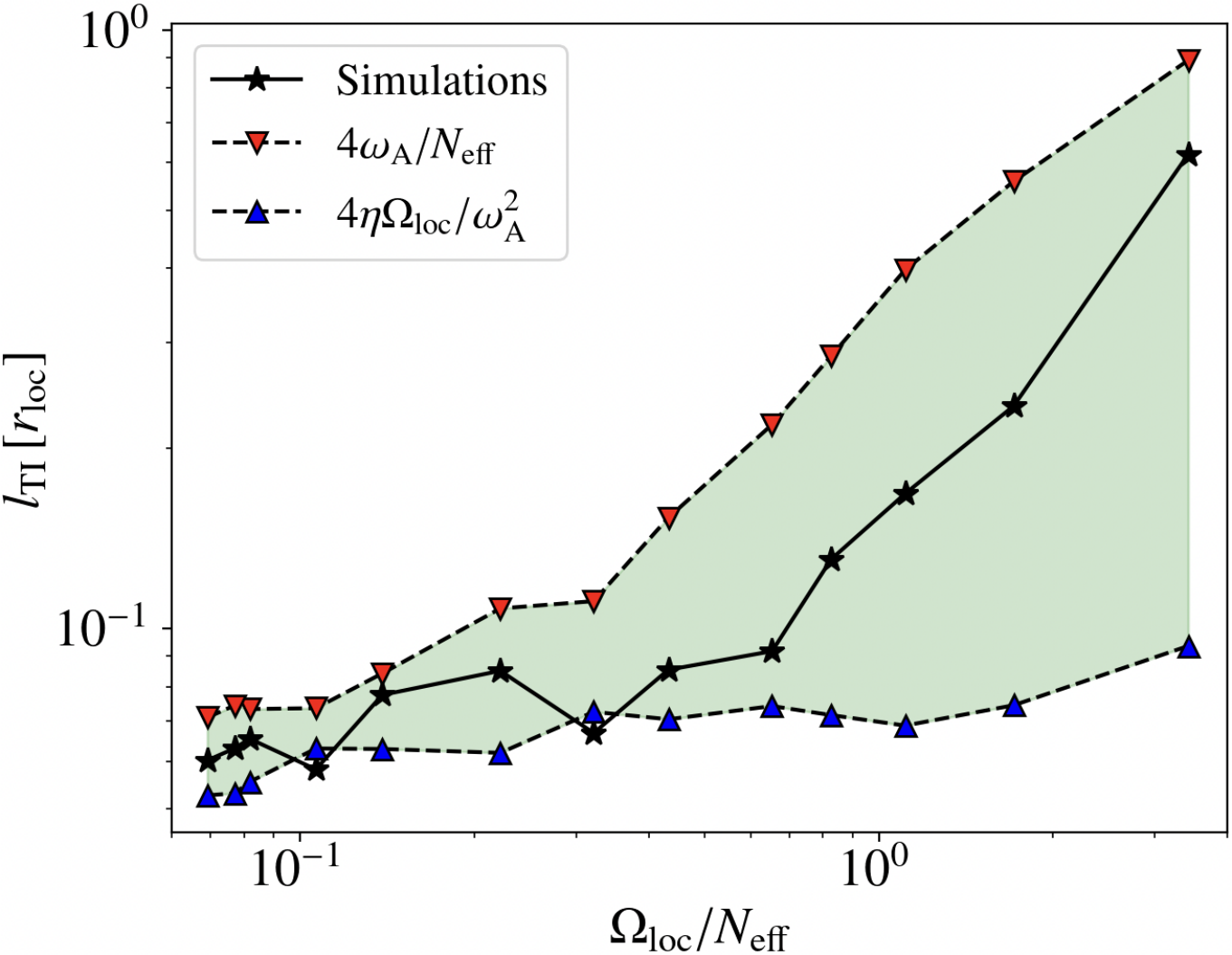}
    \caption{Length scale of the Tayler instability mode measured in the radial profile of the non-axisymmetric magnetic field (black stars) as a function of $\Omega_{\rm loc}/N_{\rm eff}$. The theoretical lower (blue triangles) and upper (red triangles) limits of the length scale are also plotted. Therefore, the region coloured in green indicates the theoretically possible length scales of the Tayler modes.}
    \label{fig:lTI}
\end{figure}

\begin{figure}
    \centering
    \includegraphics[width=\linewidth]{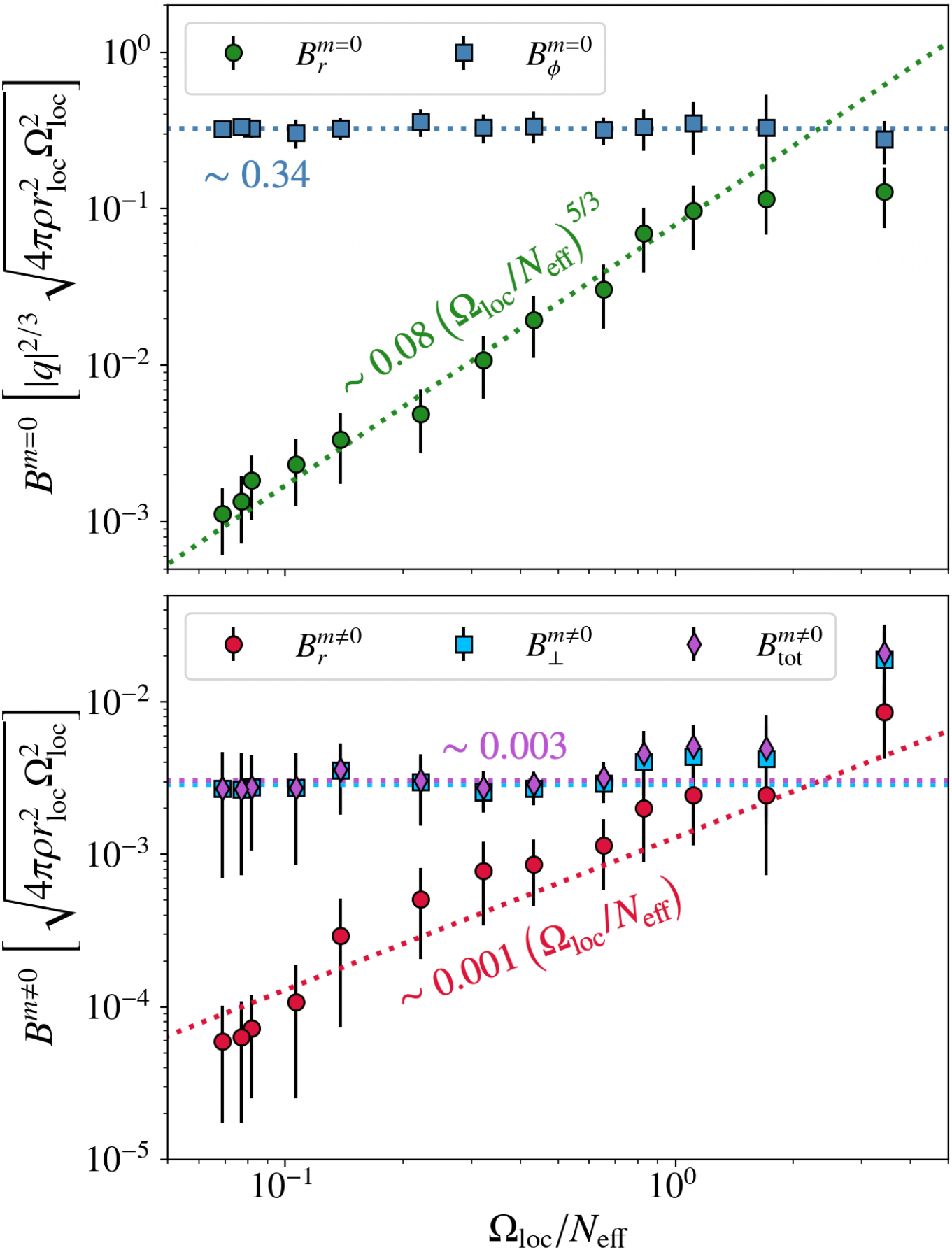}
    \caption{Top: Averaged (see Appendix~\ref{sec:average}) axisymmetric radial (green circles) and azimuthal (blue squares) as a function of $\Omega_{\rm loc}/N$. Best fitted power-laws of $\Omega_{\rm loc}/N$ are represented by the dotted lines. Bottom: Same as on top, but for the non-axisymmetric radial (red circles), perpendicular (sky blue squares), and total (purple diamonds) magnetic fields.}
    \label{fig:Ball}
\end{figure}

Fig.~\ref{fig:Ball} displays the different axisymmetric (plot on top) and non-axisymmetric (plot on bottom) components of the magnetic field as a function of $\Omega_{\rm loc}/N_{\rm eff}$. They are scaled in rotational units (see Sect.~\ref{sec:output}) and compensated by a power law of $q$ ($2/3$ and $0$ for the axisymmetric and non-axisymmetric components, respectively). The scaling laws for the axisymmetric components read:
\begin{align}\label{eq:scale_Bphi}
    B_{\phi}^{m=0} &= 0.34\sqrt{4\pi\rho r_{\rm loc}^2}|q|^{2/3}\Omega_{\rm loc}\,,\\
    \label{eq:scale_Br}
    B_r^{m=0} &= 0.08\sqrt{4\pi\rho r_{\rm loc}^2}|q|^{2/3}\Omega_{\rm loc}\left(\frac{\Omega_{\rm loc}}{N_{\rm eff}}\right)^{5/3}\,.
\end{align}
The $B_r^{m=0}$ follows the prescription derived by~\citet{fuller2019} with the prefactor $0.08$. Surprisingly, $B_{\phi}^{m=0}$ does not depend on $N_{\rm eff}$. This was not predicted by previous works, which estimated a dependence on $(\Omega_{\rm loc}/N_{\rm eff})^{1/3}$~\citep{fuller2019,barrere2025} or $(\Omega_{\rm loc}/N_{\rm eff})$~\citep{spruit2002,petitdemange2024}. This implies that the ratio between both components 
\begin{equation}\label{eq:ratio_BrB}
    \frac{B_r^{m=0}}{B_{\phi}^{m=0}}=0.24\left(\frac{\Omega_{\rm loc}}{N_{\rm eff}}\right)^{5/3} = 0.71\frac{\omega_{\rm A}}{N_{\rm eff}} \left(\frac{\Omega_{\rm loc}}{|q|N_{\rm eff}}\right)^{2/3}\,.
\end{equation}
The expected scaling law is, however, $\sim \omega_{\rm A}/N_{\rm eff}$. This expression stems from the assumed balance between the magnetic tension due to perturbations of $B_r^{m=0}$ and the magnetic pressure driving the Tayler instability, which translates into~\citep{fuller2019}
\begin{equation}
     \frac{l_{\perp}}{l_{\rm TI}}B_r^{m=0} \sim \frac{N_{\rm eff}}{\omega_{\rm A}}B_r^{m=0} \sim B_{\phi}^{m=0}\,.
\end{equation}
Therefore, this balance is not reached when the Tayler-Spruit dynamo saturates in our simulations. This discrepancy can also be observed in Fig.~\ref{fig:lTI}, where we see that $l_{\rm TI}/l_{\perp}\sim l_{\rm TI}/r_{\rm loc}\sim 4\omega_{\rm A}/N_{\rm eff}$ for $\Omega_{\rm loc}/N_{\rm eff}\in [0.07,0.4]$, while $l_{\rm TI}/r_{\rm loc}\sim 2\omega_{\rm A}/N_{\rm eff}$ for $\Omega_{\rm loc}/N_{\rm eff}\in [0.4,3]$.

For the non-axisymmetric components, we find the following relations:
\begin{align}\label{eq:scale_dBperp}
    B_{\rm tot}^{m\neq0} &\approx B_{\perp}^{m\neq0} = 0.003\sqrt{4\pi\rho r_{\rm loc}^2}\Omega_{\rm loc}\,,\\
    \label{eq:scale_dBr}
    B_r^{m\neq0} &= 0.001\sqrt{4\pi\rho r_{\rm loc}^2}\Omega_{\rm loc}\left(\frac{\Omega_{\rm loc}}{N_{\rm eff}}\right)\,.
\end{align}
Therefore, the non-axisymmetric magnetic field is largely dominated by its perpendicular component $B_{\perp}^{m\neq0}=[(B_{\theta}^{m\neq0})^2 + (B_{\phi}^{m\neq0})^2]^{1/2}$ and the ratio with the radial component follows globally well the solenoidal condition for the non-axisymmetric magnetic field
\begin{equation}
    \frac{B_r^{m\neq0}}{B_{\perp}^{m\neq0}}\approx 0.33|q|^{-2/3}\frac{\omega_{\rm A}}{N_{\rm eff}}\approx 0.43-1.5\frac{\omega_{\rm A}}{N_{\rm eff}}\,.
\end{equation}
However, \citet{fuller2019} predicted a faster decrease of both non-axisymmetric components as $\Omega_{\rm loc}/N_{\rm eff}$ decreases. Moreover, the ratio
\begin{equation}\label{eq:ratio_dBB}
    \frac{B_{\perp}^{m\neq0}}{B_{\phi}^{m=0}}= 0.026 |q|^{-4/3} \frac{\omega_{\rm A}}{\Omega_{\rm loc}} \approx 0.051-0.56\frac{\omega_{\rm A}}{\Omega_{\rm loc}}\,,
\end{equation}
can be quite small compared to the prediction $B_{\perp}^{m\neq0}/B_{\phi}^{m=0}\sim\omega_{\rm A}/\Omega_{\rm loc}$ derived by~\citet{fuller2019}.

\begin{figure}
    \centering
    \includegraphics[width=\columnwidth]{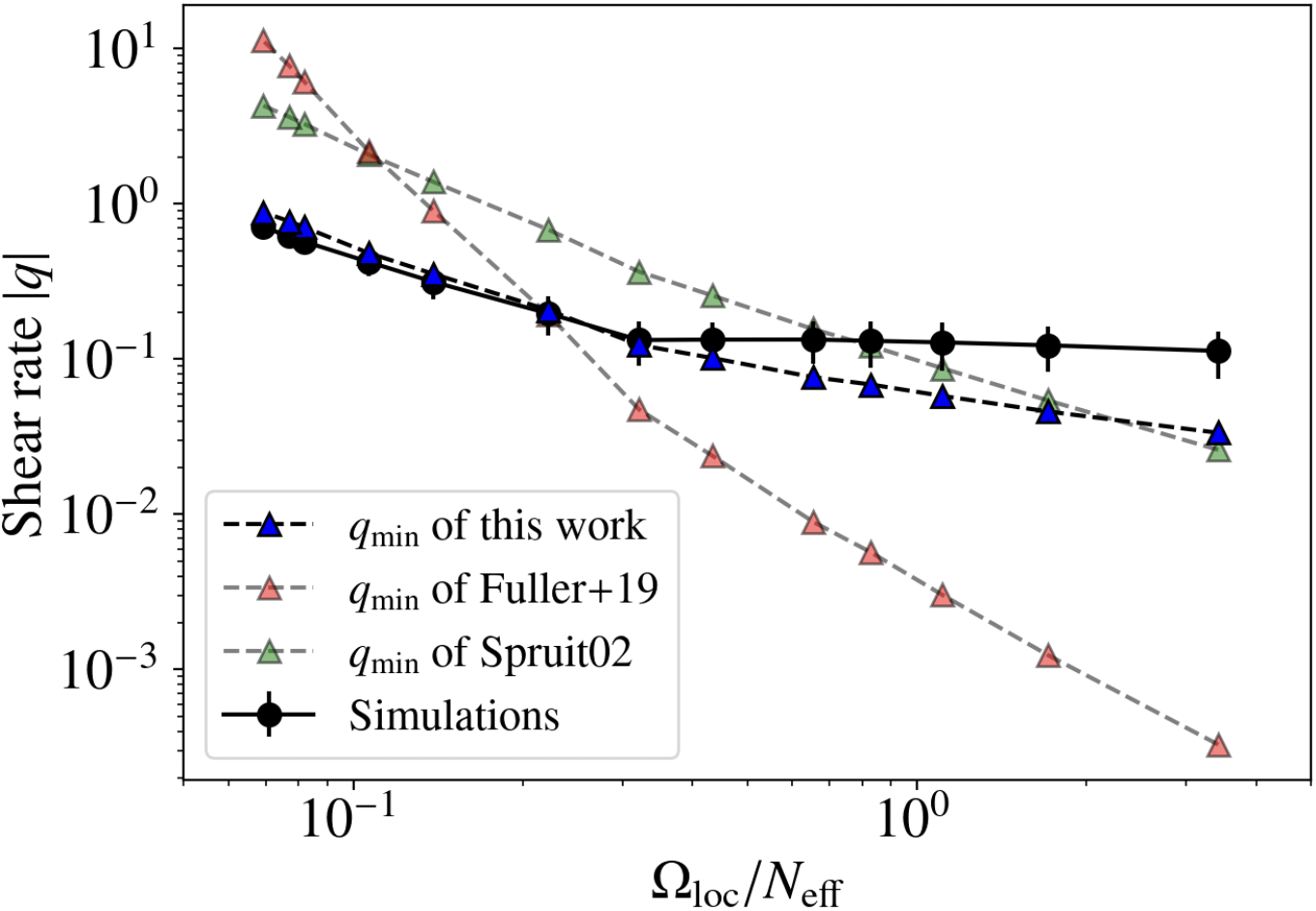}
    \caption{Different shear rates as a function of $\Omega_{\rm loc}/N$: $q$ that is measured in our simulations (black circles), minimum $q$ predicted by our scaling laws (blue triangles), by ~\citet[][red triangles]{fuller2019}, and by ~\citet[][green triangles]{spruit2002}. Note that we used a prefactor calibrated on our scaling law of $B_{\phi}^{m=0}$ for every plotted $q_{\rm min}$.}
    \label{fig:shear}
\end{figure}

Since $B_{\phi}^{m=0}$ follows Eq.~\ref{eq:scale_Bphi} and the prescription for $\omega_{\rm A,c}$ (Eq.~\ref{eq:critic}) is in global agreement with our data, we can infer a minimum shear by equating both equations:
\begin{equation}\label{eq:qMin}
    q_{\rm min} \approx 5.2\left(\frac{N_{\rm eff}}{\Omega_{\rm loc}}\right)^{3/4}\left(\frac{\eta}{r_{\rm loc}^2\Omega_{\rm loc}}\right)^{3/8}\,.
\end{equation}
This relation is plotted with the shear rate from our simulations in Fig.~\ref{fig:shear} (blue triangles and black circles, respectively). This plot confirms that our new prescription of $q_{\rm min}$ is a good lower limit for the onset of the Tayler instability in our simulations, especially at $\Omega_{\rm loc}/N_{\rm eff}\lesssim 0.4$. We also show in this figure the $q_{\rm min}$ derived by~\citet[][red triangles]{fuller2019} and~\citet[][green triangles]{spruit2002}, which predict much higher shear rates at strong stratifications ($\Omega_{\rm loc}/N_{\rm eff}\lesssim 0.25$) than the one we measured. This difference is due to the new derived scaling law for $B_{\phi}^{m=0}$ (Eq.~\ref{eq:scale_Bphi}), which does not depend on $N_{\rm eff}$.

\subsection{Different transports}\label{sec:amt_laws}

\begin{figure}
    \centering
    \includegraphics[width=\linewidth]{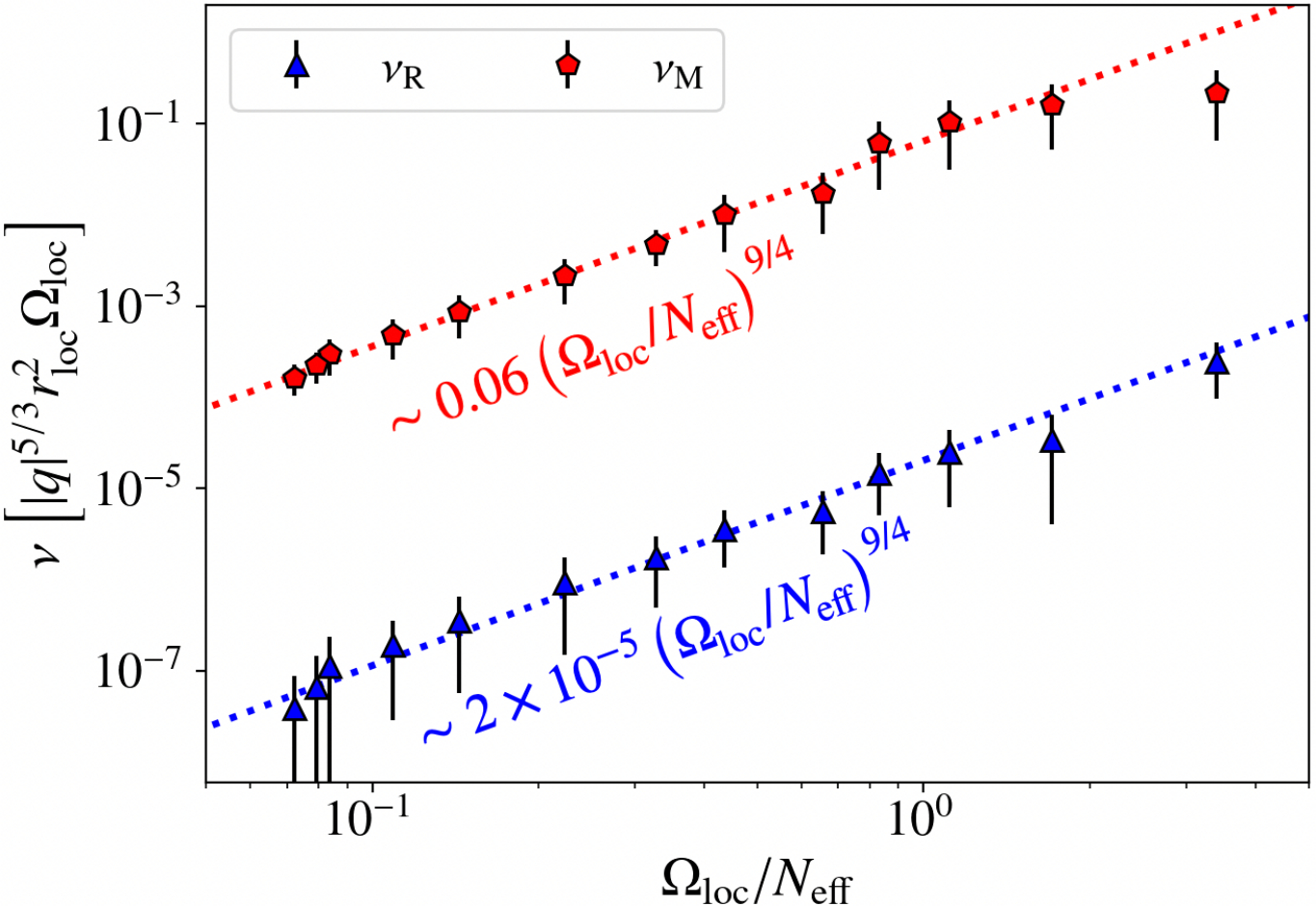}
    \caption{Same as Fig.~\ref{fig:Ball}, but for the viscosities associated with the different transport mechanisms: Reynolds stress (blue triangles) and Maxwell stress (red pentagons).}
    \label{fig:AMT}
\end{figure}

The large-scale magnetic fields generated by the Tayler-Spruit dynamo produce Maxwell stresses, which transport AM. In Fig.~\ref{fig:AMT}, we plot the viscosity associated with this transport mechanism, which follows the scaling law
\begin{equation}\label{eq:nu_M}
    \nu_{\rm M} \defeq \frac{(B_{r}B_{\phi})^{m=0}}{4\pi\rho|q|\Omega_{\rm loc}}=0.06|q|^{5/3}r_{\rm loc}^2\Omega_{\rm loc}\left(\frac{\Omega_{\rm loc}}{N_{\rm eff}}\right)^{9/4}\,.
\end{equation}
The efficiency of the transport is therefore slightly less efficient than predicted by~\citet[][$\nu_{\rm M}\propto (\Omega/N_{\rm eff})^{2}$]{fuller2019}, but much more efficient than the model of~\citet[][$\nu_{\rm M}\propto (\Omega/N_{\rm eff})^{4}$]{spruit2002}. Note that the expression we find for $\nu_{\rm M}$ does not equal the multiplication of the scaling laws for $B_{\phi}^{m=0}$ (Eq.~\ref{eq:scale_Bphi}) and $B_r^{m=0}$ (Eq.~\ref{eq:scale_Br}), which would give a more efficient transport, with $\nu_{\rm M}\propto(\Omega/N_{\rm eff})^{5/3}$. The explanation of this difference relies on the separate latitudinal locations of both components, which are at the colatitude $\theta\approx45^{\circ}$ and close to the polar axis for $B_{\phi}^{m=0}$ and $B_r^{m=0}$, respectively. This could not be captured by previous theoretical models because of their one-zone character.

Besides, as seen in Fig.~\ref{fig:AMT}, Maxwell stresses dominate the AM transport driven by flow turbulence, which is quantified by the Reynolds stresses
\begin{equation}
\begin{split}
    \nu_{\rm R} \defeq \frac{v_{r}^{m\neq0}v_{\phi}^{m\neq0}}{|q|\Omega_{\rm loc}}&=\SI{2e-5}{}|q|^{5/3}r_{\rm loc}^2\Omega_{\rm loc}\left(\frac{\Omega_{\rm loc}}{N_{\rm eff}}\right)^{9/4}\\
    &\approx\SI{3e-4}{}\nu_{\rm M}\,.
\end{split}
\end{equation}

$\nu_{\rm R}$ significantly differs from the estimates of~\citet{fuller2019}, by the factor $|q|^{-2/3}(\Omega_{\rm loc}/N_{\rm eff})^{1.1}$, if we ignore the prefactor value. The discrepancy is certainly related to the difficulty to estimate the turbulent velocity for one-zone models. \citet{fuller2019} estimated the non-axisymmetric velocities by using the incompressibility condition and the assumption of quasi-magnetogeostrophic balance to link the turbulent magnetic fields and velocities 
\begin{equation}
    v_{\perp}^{m\neq0}\sim\frac{N_{\rm eff}}{\omega_{\rm A}}v_r^{m\neq0}\sim \frac{\omega_{\rm A}}{\Omega_{\rm loc}} v_{\rm A, \perp}^{m\neq0}\,,
\end{equation}
where $v_{\rm A, \perp}^{m\neq0}=B_{\rm \perp}^{m\neq0}/\sqrt{4\pi\rho}$. Therefore, the difference may be a consequence of the tension we noticed for the scaling law of $B_{\rm \perp}^{m\neq0}$ (Eq.~\ref{eq:scale_dBperp}).

\section{Link with observations}\label{sec:applications}
The two previous sections provide a new numerical analysis and results on the Tayler-Spruit dynamo. This fosters the following section, where we confront our results to the observations. We first discuss the impact of the dynamo-generated magnetic field on the asteroseismic properties to give constraints on the signal (Sect.~\ref{sec:astero}). Second, we glimpse how our new prescriptions for the different transports affect stellar evolution (Sect.~\ref{sec:evol}).

\subsection{Asteroseismic detection of the magnetic field}\label{sec:astero}
The recent asteroseismic observations of red giants provide the first observational constraints on the average radial magnetic field, usually noted $\sqrt{\langle B_r^2 \rangle}$~\citep{li2022,li2023,hatt2024}. The magnetic shift parameter $\delta\nu_{\rm mag}$ gives an estimate of $\sqrt{\langle B_r^2 \rangle}$ and is estimated by fitting the asteroseismic data. Observational studies find that the detected radial field located in the helium-burning shell (HBS) is $\sqrt{\langle B_r^2 \rangle}\sim 10^4-\SI{e5}{G}$. According to our scaling laws for the Tayler-Spruit dynamo, we can also estimate the generated radial magnetic field 
\begin{equation}
    B_r^{m=0}\approx \SI{e-3}{}\left(\frac{\Omega_{\rm HBS}}{\SI{0.7}{\mu Hz}}\right)^{8/3}\left(\frac{N_{\rm HBS}}{\SI{2e4}{\mu Hz}}\right)^{-5/3}\SI{}{G}\,, 
\end{equation}
for $|q|\sim 1$ and parameters relevant for a HBS: $r_{\rm HBS}=\SI{0.03}{R_{\odot}}$, and $\rho_{\rm HBS}=\SI{0.01}{g.cm^{-3}}$. As already expected by~\citet{li2022}, the Tayler-Spruit dynamo cannot explain the observed field strengths.

Moreover, the usual expression used for $\delta\nu_{\rm mag}$ assumes a radial magnetic field not too weak compared to the horizontal components. However, in an HBS, the Tayler-Spruit dynamo maintains a strong toroidal field 
\begin{equation}
    B_{\phi}^{m=0}\approx \SI{e5}{}\left(\frac{\Omega_{\rm HBS}}{\SI{0.7}{\mu Hz}}\right)\SI{}{G}\approx \SI{e8}{}B_{r}^{m=0}\,.
\end{equation}
The usual expression of the $\delta\nu_{\rm mag}$ is therefore not relevant for the magnetic field produced by the Tayler-Spruit dynamo. According to~\citet{li2022}, the main change for a strong toroidal field is the variation of $\delta\nu_{\rm mag}$ with the frequency of the oscillation spectra, which becomes $\delta\nu_{\rm mag}\propto \nu^{-1}$, instead of $\delta\nu_{\rm mag}\propto \nu^{-3}$ for a strong radial component. So far, the assumption of a strong azimuthal field is not consistent with any of the $\delta\nu_{\rm mag}$ measured in red giants. Note also that the non-axisymmetric magnetic field is not negligible $B_{\perp}^{m=0}\approx\SI{e4}{G}\approx 0.1B_{\phi}^{m=0}$, which may also have an effect on the signal.

\begin{figure}
    \centering
    \includegraphics[width=\linewidth]{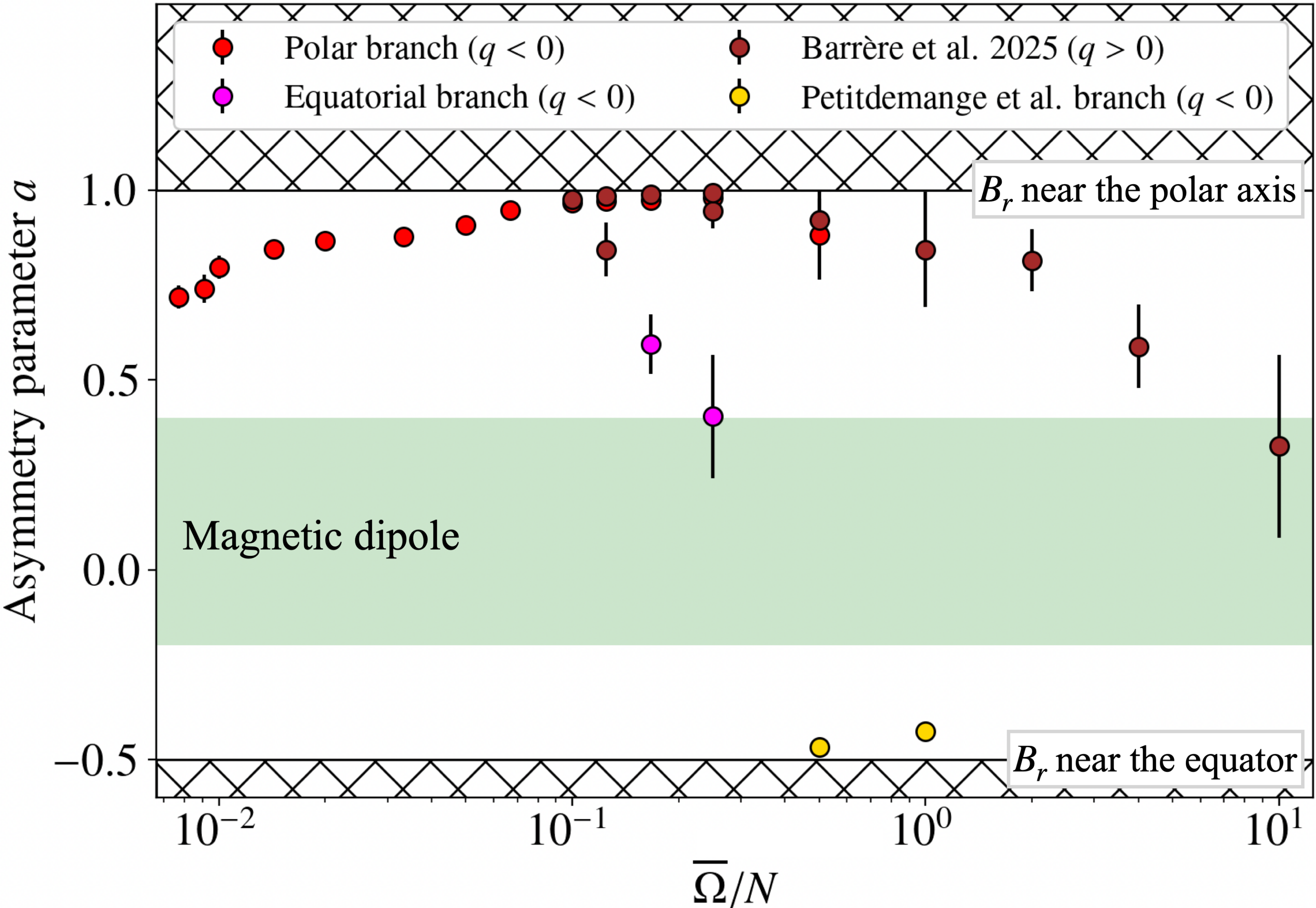}
    \caption{Relation between the asymmetry parameter $a$ and the ratio $\overline{\Omega}/N$ for different sets of simulations: the polar (red) and equatorial (magenta) dynamos of this paper, the stratified simulations for proto-magnetars of~\citet[][brown]{barrere2025}, and the reproduction of two runs from~\citet[][yellow]{petitdemange2023}. The green region indicates the range of asymmetry parameters $[-0.2,0.4]$, which can be explained by a pure magnetic dipole.}
    \label{fig:asym}
\end{figure}

Despite the disagreement between the magnetic fields generated by the Tayler-Spruit dynamo and the fields detected in red giant HBS, constraining other properties of the dynamo-generated magnetic field is crucial to better interpret future magnetic field observation. In particular, the large-scale topology of $\sqrt{\langle B_r^2 \rangle}$ can be constrained by using the dimensionless asymmetry parameter $a$. It characterises the latitudinal distribution of $B_r$ in the oscillation cavity, which is weighted by the second-degree Legendre polynomial $P_2$~\citep{li2022,mathis2023}:
\begin{equation}
    a \defeq\frac{\int^{r_o}_{r_i}K(r)\int\int_{S}B_r^2P_2(\cos\theta)d\Omega dr}{\int^{r_o}_{r_i}K(r)\int\int_SB_r^2 d\Omega dr}\in[-0.5,1]\,,
\end{equation}
where $S$ is the spherical surface, $r_i$ and $r_o$ are the inner and outer radii of the oscillation cavity, and 
\begin{equation}
    K(r) \defeq \frac{\rho^{-1}(N/r)^3}{\int^{r_o}_{r_i}\rho^{-1}(N/r)^3dr}\,,
\end{equation}
is a weighted function depending on the stratification profile in the radiative zone. $a$ can span the range $[-0.5,1]$, whose lower and upper limits describe a $B_r$ near the equator or the polar axis, respectively. The values of $a$ in our simulations are gathered in Fig.~\ref{fig:asym}, where we also plotted the value of $a$ for all simulations from~\citet{barrere2025} and two reproduced from~\citet{petitdemange2024}. On the one hand, the equatorial Tayler-Spruit dynamo from~\citet{petitdemange2024} has values of $a$ close to $-0.5$, but our equatorial dynamo shows much higher time-averaged values ($0.4$, $0.59$). The latter values are certainly overestimated because the measure is polluted by the remnant of $B_r$ produced by a transient close to the polar axis at the beginning of the simulation. For both simulations, the value of $a$ reaches $\sim0.1$ at the last timesteps, but much longer integration times would certainly be required to reach a converged value. On the other hand, our polar dynamo is in continuity with less stratified simulations of~\citet{barrere2025}, with $a\in[0.72,0.97]$. On the observation side, $a$ can span the whole range of possible values~\citep{hatt2024}, suggesting a wide variety of magnetic field geometries, and so different formation mechanisms. Therefore, the Tayler-Spruit dynamo may produce the magnetic fields in stars with high $a$.

Finally, the impact of magnetic fields on the magneto-gravito-inertial (MGI) modes propagating in fast-rotating main-sequence stars ($\gamma$ Dor or slow pulsating B stars) may also be detected in the near future thanks to adapted asteroseismic diagnostics~\citep[e.g.][]{dhouib2022,lignieres2024}. Near the bottom of the radiative zone in a $\gamma$ Dor (where $r_{\rm Dor}=\SI{0.34}{R_{\odot}}$, and $\rho_{\rm Dor}=10^{2}\SI{}{g.cm^{-3}}$), the Tayler-Spruit dynamo would produce large-scale magnetic fields with the strengths 
\begin{align}
    B_{\phi}^{m=0}&\approx \SI{e6}{}\left(\frac{\Omega_{\rm Dor}}{\SI{10}{\mu Hz}}\right)\SI{}{G}\,,\\
    B_{r}^{m=0}&\approx \SI{e4}{}\left(\frac{\Omega_{\rm Dor}}{\SI{10}{\mu Hz}}\right)^{8/3}\left(\frac{N_{\rm Dor}}{\SI{300}{\mu Hz}}\right)^{-5/3}\SI{}{G}\approx 10^{-2} B_{\phi}^{m=0}\,.
\end{align}
The radial magnetic field is therefore much stronger than in the HBS of red giants. It is also close enough to $B_{\phi}^{m=0}$ for the  detection method developed by~\citet{lignieres2024} to remain relevant. Note that the presence of the strong $B_r^{m=0}$ may also cause, at least, a partial suppression of the MGI  modes~\citep{rui2023,barrault2025}. Moreover,~\citet{dhouib2022} show that intense $B_{\phi}^{m=0}\sim\SI{e5}{G}$ at the equator could be detectable. Therefore, despite $B_{\phi}^{m=0}$ tending to $0$ at the equator, it may be strong enough around the equator to still significantly affect MGI modes. The azimuthal field on the equatorial dynamo might, however, be more easily detectable if it can be maintained for stronger stratifications.

\subsection{Stellar internal rotation}\label{sec:evol}
Our analysis proposes new 1D prescriptions for (i) the minimum shear rate required to trigger the Tayler-Spruit dynamo (see Sect.~\ref{sec:mag_laws}) and (ii) the AM transport of AM (see Sect.~\ref{sec:amt_laws}). Once implemented in a stellar evolution code, they may change the rotation and chemical abundances obtained in previous 1D evolution studies that include the Tayler-Spruit dynamo. Here, we therefore attempt to foresee whether the dynamo can explain the measured internal rotation rates. 

Assuming $q\sim 1$, our expression of $\nu_{\rm M}$ (Eq.~\ref{eq:nu_M}) is not far from the prescription derived by~\citet{fuller2019}. We can then fit the latter
\begin{equation}
    \nu_{\rm M} = \alpha^3 r_{\rm loc}^2\Omega_{\rm loc}\left(\frac{\Omega_{\rm loc}}{N_{\rm eff}}\right)^{2}\,,
\end{equation}
to our data in order to estimate a prefactor $\alpha$ calibrated to our simulations. We obtain $\alpha\approx0.36$, which is close to the values calibrated to the near-core rotation of red giants by~\citep[][$\alpha\approx1$ when the dynamo operates at $q>q_{\rm min}$]{fuller2019} and~\citep[][$\alpha\approx0.25$ when $q_{\rm min}$ is ignored]{fuller2022}. Our value of $\alpha$ is close to the appropriate calibration estimated by~\citet[][]{eggenberger2019b} to reproduce the near-core rotation of subgiants ($\alpha\approx0.5$) but slightly insufficient for red giants ($\alpha\approx1.5$), according to them. Nonetheless, for the very strong stratifications of the HBS, the minimum shear we predict (Eq.~\ref{eq:qMin})
\begin{equation}
    q_{\rm min}\approx0.67\left(\frac{\Omega_{\rm HBS}}{\SI{0.7}{\mu Hz}}\right)^{-9/8}\left(\frac{N_{\rm HBS}}{\SI{2e4}{\mu Hz}}\right)^{3/4}\left(\frac{\eta}{\SI{100}{cm^2.s^{-1}}}\right)^{3/8}\,,
\end{equation}
is much smaller than the estimate of~\citet[][with $\alpha=1$]{fuller2019} 
\begin{equation}
    q_{\rm min}\approx 472\left(\frac{\Omega_{\rm HBS}}{\SI{0.7}{\mu Hz}}\right)^{-13/4}\left(\frac{N_{\rm HBS}}{\SI{2e4}{\mu Hz}}\right)^{5/2}\left(\frac{\eta}{\SI{100}{cm^2.s^{-1}}}\right)^{3/4}\,.
\end{equation}
Therefore, the dynamo may operate in a larger radiative region, such that the transport is efficient enough to match the red giant internal rotation. Note that the same conclusion can be drawn by using the general formulation for the transport developed by~\citet{eggenberger2022b}:
\begin{equation}
    \nu_{\rm M} = \frac{\Omega_{\rm loc}r_{\rm loc}}{|q|}\left(C_{\rm T}|q|\frac{\Omega_{\rm loc}}{N_{\rm eff}}\right)^{3/n}\left(\frac{\Omega_{\rm loc}}{N_{\rm eff}}\right)\,,
\end{equation}
where $n=1$ or $n=3$ to obtain the prescriptions of~\citet{spruit2002} or~\citet{fuller2019}, and $C_{\rm T}$ is a calibrating prefactor. Once fitted to our simulations, we can calibrate $n\approx2.4$ and $C_{\rm T}\approx0.1$.

For main-sequence intermediate-mass stars,~\citet{moyano2023} showed that the Tayler-Spruit dynamo as originally modelled by~\citet{spruit2002} can explain the observed uniform rotation in the radiative zone in $\gamma$ Dor~\citep{vanreeth2018}. Therefore, our Tayler-Spruit dynamo transports largely enough AM to reproduce the rotation of these stars. 

Finally, a few evolution models include a magnetic torque-induced transport for massive stars. Most of them use the original scaling laws~\citep{heger2005,maeder2014,wheeler2015,aguilera2018,griffiths2022}, while only~\citet{fuller2022} implemented those from~\citet{fuller2019}.
An interesting indication of efficient AM transport is the rotation period of the remaining compact object after the supernova explosion, especially neutron stars (NS)~\citep[][$P\sim\SI{100}{ms}$]{igoshev2022}. Assuming no braking or spin-up mechanisms during and after the explosion, \citet{heger2005} show that the original Tayler-Spruit dynamo produces NS rotation periods ten times smaller than those measured by observations. This makes our Tayler-Spruit dynamo a promising candidate to spin down the progenitor core efficiently and better match the observations, but this is still to be confirmed or not by future evolution models of magnetised massive stars.  

\section{Limits of the simulations}\label{sec:discussion}
\subsection{Viscosity and diffusivities}
Like most numerical simulations modelling astrophysical objects, our models use unrealistic viscosities and diffusivities, which are several orders of magnitude too high. In the case of low-mass stars, we can estimate that the (molecular or radiative) kinematic viscosity $\nu$ and the resistivity $\eta$ reach a maximum of $\sim\SI{e3}{cm^2.s^{-1}}$ in the radiative zone, during the red giant phase~\citep{rudiger2015}. This implies a maximum Ekman number in the HBS
\begin{equation}
    E \approx \SI{5e-11}{}\left(\frac{\nu}{\SI{e3}{cm^2.s^{-1}}}\right)\left(\frac{\Omega_{\rm HBS}}{\SI{0.7}{\mu Hz}}\right)^{-1}\,,
\end{equation}
and, assuming a thermal diffusivity around $\kappa\sim\SI{e8}{cm^2.s^{-1}}$~\citep{garaud2015}, the Rayleigh number reads
\begin{equation}
    Ra\approx \SI{3e24}{}\left(\frac{N_{\rm HBS}}{\SI{2e4}{\mu Hz}}\right)^2\left(\frac{\nu}{\SI{e3}{cm^2.s^{-1}}}\right)^{-1}\left(\frac{\kappa}{\SI{e8}{cm^2.s^{-1}}}\right)^{-1}\,.
\end{equation}
Therefore, $E$ and $Ra$ in stars are respectively at least 6 orders of magnitude smaller and 13 orders of magnitude larger than in our simulations. While they could be pushed to $E\gtrsim 10^{-7}$ and $Ra\lesssim 10^{13}$ with the MagIC code, the realistic values remain far beyond the capacity of any modern supercomputers. We could expect the development of the dynamo to be favoured in the realistic regime, as the classical and magnetic Reynolds numbers are very large; that is, the viscous and resistive times are much longer than the advective time, in stars ($\sim10^{12}$ and $\sim10^9$ in the Sun, respectively). However, this regime is still very poorly understood, so the extrapolation of our scaling laws to this regime remains an open question.

\subsection{Rotation profile}
We chose to impose a shellular rotation profile (Eq.~\ref{eq:rotation}) for the body force, which maintains the differential rotation. This profile is justified because the fluid is in a viscous regime with $Pr(N/\overline{\Omega})^2>1$ for every simulation, except the less stratified one ($N/\overline{\Omega}=2$). However, for post-main sequence low- or intermediate-mass stars, a part of the radiative zone contracts quickly, which influences the structure of the flow. \citet{gouhier2021,gouhier2022} show that the contraction creates radial differential rotation, but also a latitudinal component in the presence of large-scale magnetic fields and despite a strong stratification ($Pr(N/\overline{\Omega})^2=10^4$). The latter component would favour the development of MRI if the poloidal component is not too strong. Therefore, if the radiative zone has a strong magnetic field at the beginning of the subgiant phase, our rotation profile may not be relevant for subgiants.

\subsection{Boussinesq approximation}
In these simulations, we assumed that the fluid follows the Boussinesq approximation, that is, the variations of the fluid density are neglected except in the buoyancy term. This approximation is practical because it significantly reduces the numerical cost by filtering out the acoustic waves and simplifying the MHD equations. Nonetheless, this approximation implies a uniform density profile, which is unrealistic, as shown by stellar evolution models. Moreover, the density gradient steepens near the core as it contracts during the evolution. The effect of these gradients on the Tayler-Spruit dynamo has never been investigated yet. Therefore, the anelastic approximation with polytropic or realistic density profiles from evolution models will be used in future work to study the Tayler-Spruit dynamo at different evolution stages.

\section{Conclusions}\label{sec:conclusions}

In this paper, we investigated the Tayler-Spruit dynamo in the context of stellar physics using 3D direct numerical simulations. Like in previous studies~\citep[e.g.][]{meduri2024}, we use a volumetric forcing of the differential rotation to notably avoid triggering instabilities caused by a spherical Taylor-Couette configuration. We demonstrated the existence of a Tayler-Spruit dynamo in the shear flow of a stellar radiative zone, whose turbulence is clearly driven by a Tayler instability near the polar axis. Another novelty is the coexistence of the Tayler-Spruit dynamo with a dynamo developing around the equatorial plane and driven by an instability sharing properties of both azimuthal MRI and Tayler instability. While the equatorial dynamo only operates in a weakly stratified regime ($\overline{\Omega}/N\geq\SI{0.11}{}$), the Tayler-Spruit dynamo is maintained for very strong stratifications ($\overline{\Omega}/N\geq\SI{7.7e-3}{}$). After quantifying the effect of strong stratification on Tayler modes, we inferred scaling laws calibrated on our simulations for the magnetic fields and the different transports. We can summarise our results in two main conclusions:
\begin{itemize}
    \item We find that the radial magnetic field $B_r^{m=0}$ follows the scaling derived by~\citet{fuller2019}, but the azimuthal component $B_{\phi}^{m=0}$ does not depend on $N_{\rm eff}$ (Eq.~\ref{eq:scale_Bphi}). Therefore, $B_{\phi}^{m=0}$ remains stronger in strongly stratified regimes than previously foreseen. Since we also validate the usual expression of the critical magnetic field (Eq.~\ref{eq:critic}) to activate the Tayler instability, we infer a new prescription for the minimum shear $q_{\rm min}$ (Eq.~\ref{eq:qMin}) from the new scaling law of $B_{\phi}^{m=0}$. This new expression predicts that weak shear rates can trigger the Tayler-Spruit dynamo even in strongly stratified fluids.
    \item We confirm that the AM transport is dominated by the large-scale magnetic fields. The scaling law we fitted shows that the transport is slightly less efficient than predicted by~\citet{fuller2019}. However, $q_{\rm min}$ can be much smaller than analytically predicted, which suggests that the dynamo should operate in larger regions, and so extract more AM. The AM transport by the turbulent flow is less efficient by a factor $\approx\SI{3.3e3}{}$. 
\end{itemize}
Our analysis therefore provides important new quantitative predictions for the magnetic fields and transports generated by the Tayler-Spruit dynamo in stably stratified fluids. This finding, however, makes the physics behind the Tayler-Spruit dynamo more complex and raises new theoretical challenges.

An important question is how the discrepancies with the previous theories can be overcome. Equation~\ref{eq:ratio_dBB} suggests that the saturation of the Tayler instability in our simulations is consistent with dissipation by an Alfvénic cascade proposed by~\citet{fuller2019}. However, the scaling law of $B_{\phi}^{m=0}$ (Eq.~\ref{eq:scale_Bphi}) and the expression of the ratio $B_r^{m=0}/B_{\phi}^{m=0}$ (Eq.~\ref{eq:ratio_BrB}) show a tension with analytical studies for the saturation of large-scale magnetic fields. A first explanation could rely on the latitudinal average we do to measure the different fields, because it does not take into account the fact that the maximum of $B_{\phi}^{m=0}$ and $B_{r}^{m=0}$ are located at different latitudes. However, after measuring the ratio $B_{r}^{m=0}/B_{\phi}^{m=0}$ at the colatitude where $B_{r}^{m=0}$ is maximum ($\theta\sim 2^{\circ}$), we still do not find the expected scaling law (see Appendix~\ref{sec:BrBphi_loc}) and the value of $B_{\phi}^{m=0}$ is uncertain as it tends towards 0 in this region. Therefore, this tension may be related to an inaccurate estimate of the large-scale magnetic field dissipation rates in the theory of \citet{fuller2019}. Our numerical study thus fosters the derivation of a revised analytical model to explain the scaling laws we have determined.

As evoked in Sect.~\ref{sec:bistable}, our polar dynamo shows several similarities with the `strong dipolar' dynamo reported by~\citet{barrere2023}, which operates when $q>0$. Indeed, the geometry of the generated large-scale magnetic fields is the same. When $q>0$, the Tayler modes also develop near the polar axis, but are also located closer and closer to the inner spherical boundary as $\overline{\Omega}/N$ decreases. This difference can be explained by the forcing of differential rotation, which consisted in imposing fixed different rotation rates on both boundaries. This method tends to produce strong shear near the inner boundary, which favours the development of the dynamo. Therefore, this difference is unlikely to be related to the sign of $q$. Also, unlike our polar dynamo, the Tayler-Spruit dynamo at $q>0$ is difficult to maintain for strong stratifications, even at $Pm=4$~\citep{barrere2025}, but this limit should be tested using our volumetric forcing in future simulations. A comparison of the magnetic field strengths, and so the scaling laws, between the dynamos is not straightforward, because \citet{barrere2025} use volume averages of energies associated with different magnetic field components to estimate magnetic strengths. Besides, they keep the poloidal/toroidal decomposition instead of the decomposition according to the spherical coordinates. This choice was justified as they compared their results to the poorly constrained magnetic fields of magnetars. Nonetheless, their scaling laws are in global agreement with the predictions of~\citet{fuller2019} but with a normalisation factor of $\alpha \approx0.01$. Therefore, the magnetic fields are weaker than those generated by the Tayler-Spruit dynamo studied in this paper.

A surprising result that was not predicted by previous analytical studies is the bistability of two dynamos. This situation was also presented by~\citet{barrere2023}, who reported a bistability between two Tayler-Spruit dynamos differing from the intensity and the equatorial symmetry of the generated fields: strong and dipolar on the one hand, and weak and hemispherical on the other hand. \citet{barrere2026a} shows that the latter dynamo quickly disappears as the magnetic field branches off to the strong dynamo for $\overline{\Omega}/N\lesssim 4$. However, here, the magnetic field does not seem to branch off to the polar dynamo when the equatorial dynamo cannot be maintained ($\overline{\Omega}/N\lesssim 0.125$). Since both dynamos were obtained using initial poloidal magnetic fields with opposite equatorial symmetries, a parametric study varying the ratio of between the energies associated to the initial symmetric and antisymmetric components of the poloidal field could enable the investigation of the transition between both dynamos. The breaking of the flow equatorial symmetry could also play a role in the transition, and magnetic reversals could also emerge, as observed for the Tayler-Spruit dynamo with $q>0$~\citep{barrere2026a} and convective flows~\citep{gissinger2012}.

A remaining crucial question is the subcritical transition to the Tayler-Spruit dynamo. In stellar evolution models including magnetic effects, the dynamo is assumed to operate when the shear rate exceeds the threshold $q_{\rm min}$. This criterion is very simplistic because it cannot grasp the highly nonlinear mechanism enabling the subcritical transition to a dynamo state. For instance, \citet{riols2013} invoked global homoclinic and heteroclinic bifurcations to explain the transition to the MRI-driven dynamo in shearing boxes. A key ingredient is the minimal seed, that is, the weakest magnetic field with the right finite-amplitude disturbances that attracts to the dynamo branch. They act as edge states separating the non-dynamo from the dynamo states. Recent methods have been developed to identify these seeds~\citep{mannix2022}, and applied to the geomagnetic dynamo~\citep{skene2024}.

Finally, as discussed in Sect.~\ref{sec:applications}, our numerical investigation has significant implications to (i) determine the impact of the Tayler-Spruit dynamo on the asteroseismic signal, and (ii) explain the inner rotation and surface abundances in stars. First, our study confirms that the Tayler-Spruit dynamo cannot explain the radial field observed in red giants~\citep[as suggested by][]{li2022}. However, it fosters the search for asteroseismic signals impacted by strong magnetic fields dominated by their azimuthal component. In particular, this would imply a magnetic shift proportional to the inverse of the frequency in the oscillation spectrum. In $\gamma$ Dor, the azimuthal magnetic field could be detectable~\citep{dhouib2022}, but the strong radial component may also partially suppress the MGI modes~\citep{rui2023,barrault2025}. Future models of propagating gravity modes including magnetic field configurations stemming from numerical models of the Tayler-Spruit dynamo would clarify the impact on the asteroseismic signal. Second, considering previous evolution models, the efficient transport should be enough to explain the rotation of red giants. Nonetheless, it remains uncertain whether the dynamo helps to reproduce the rotation of subgiants, which requires less efficient transport. An alternative scenario related to the strong contraction of the core may be more likely, as suggested by the work of~\citet{gouhier2021,gouhier2022}. To determine in which evolution stages the Tayler-Spruit dynamo can explain the observed internal rotations, future grids of stellar evolution models should include our prescriptions of $\nu_{\rm M}$ (Eq.~\ref{eq:nu_M}) and $q_{\rm min}$ (Eq.~\ref{eq:qMin}). Finally, estimating the chemical mixing of elements would be very interesting to test whether the Tayler-Spruit dynamo can explain the observed surface abundances of chemical elements. The low values of the Reynolds stress and the small radial length scale of the Tayler instability at strong stratifications suggest weak mixing, as predicted by analytical studies~\citep{spruit2002,fuller2019}. The measure of the mixing would, however, require methods more robust than the analytical proxy used in these studies, such as simulating the advection of passive scalars~\citep[similarly to][]{rincon2025}.

\begin{acknowledgements}
      We thank the referee for his/her/their thorough reading and comments, which have been very useful to improve the manuscript. PB and ARS thank J. Guilet and R. Raynaud for fruitful discussions. PB, PE, CR, and MM acknowledge support from the SNF grant No 219745 (Asteroseismology of transport processes for the evolution of stars and planets). Numerical simulations have been carried out at the CINES on the Jean-Zay supercomputer and at the TGCC on the supercomputer IRENE-ROME (DARI project A0170410317).
\end{acknowledgements}

\bibliographystyle{aa}
\bibliography{biblio}

\begin{appendix} 
\section{Variation of the relaxation timescale for the volumetric forcing}\label{sec:forcingTime}
\begin{figure*}[h!]
    \sidecaption
    \centering
    \includegraphics[width=12cm]{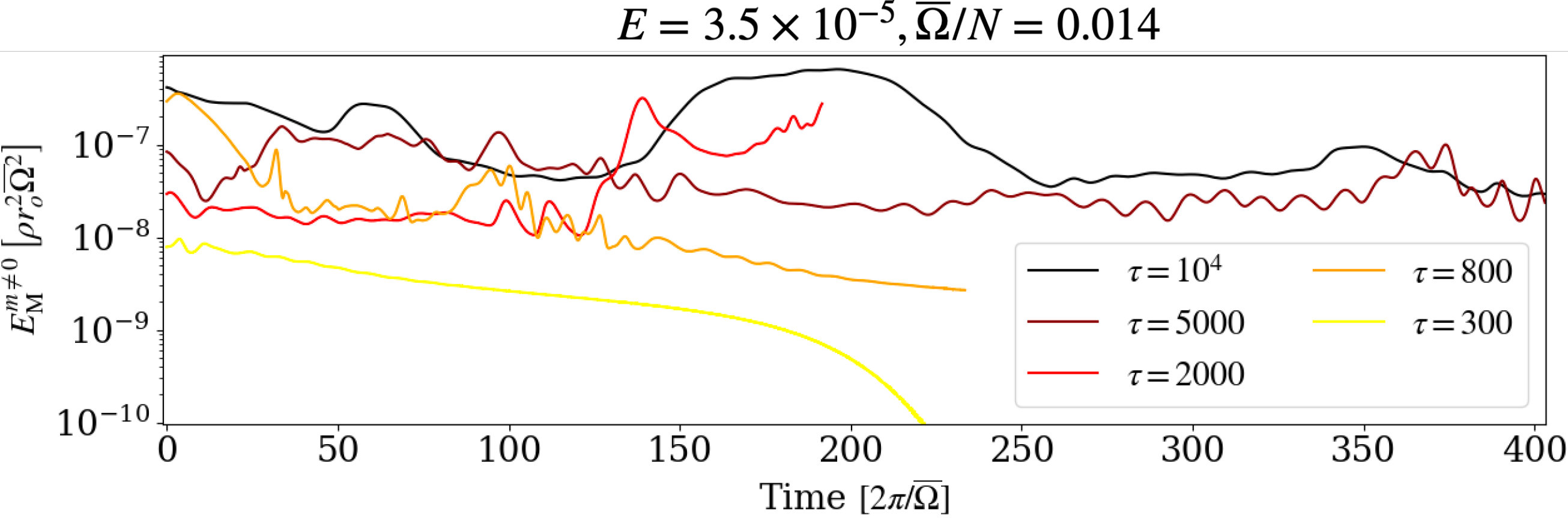}
    \caption{Time series of the non-axisymmetric magnetic energy for several values of the relaxation time scale used for the volumetric forcing.}
    \label{fig:forcingTime}
\end{figure*}

The time series displayed in Fig.~\ref{fig:forcingTime} show that the turbulence created by the Tayler-Spruit dynamo is damped when the relaxation timescale $\tau^{-1}\approx \SI{e-3}{} \approx 29\times E$ (viscous units).

\section{Magnetic $m$-spectra}\label{sec:spectra}
The magnetic spectra show the presence of significant large-scale axisymmetric ($m=0$) poloidal and toroidal fields produced by the Tayler-Spruit dynamo. We clearly see that the dominant non-axisymmetric mode is $m=1$, which is compatible with the Tayler instability.
\begin{figure*}[h!]
    \sidecaption
    \centering
    \includegraphics[width=12cm]{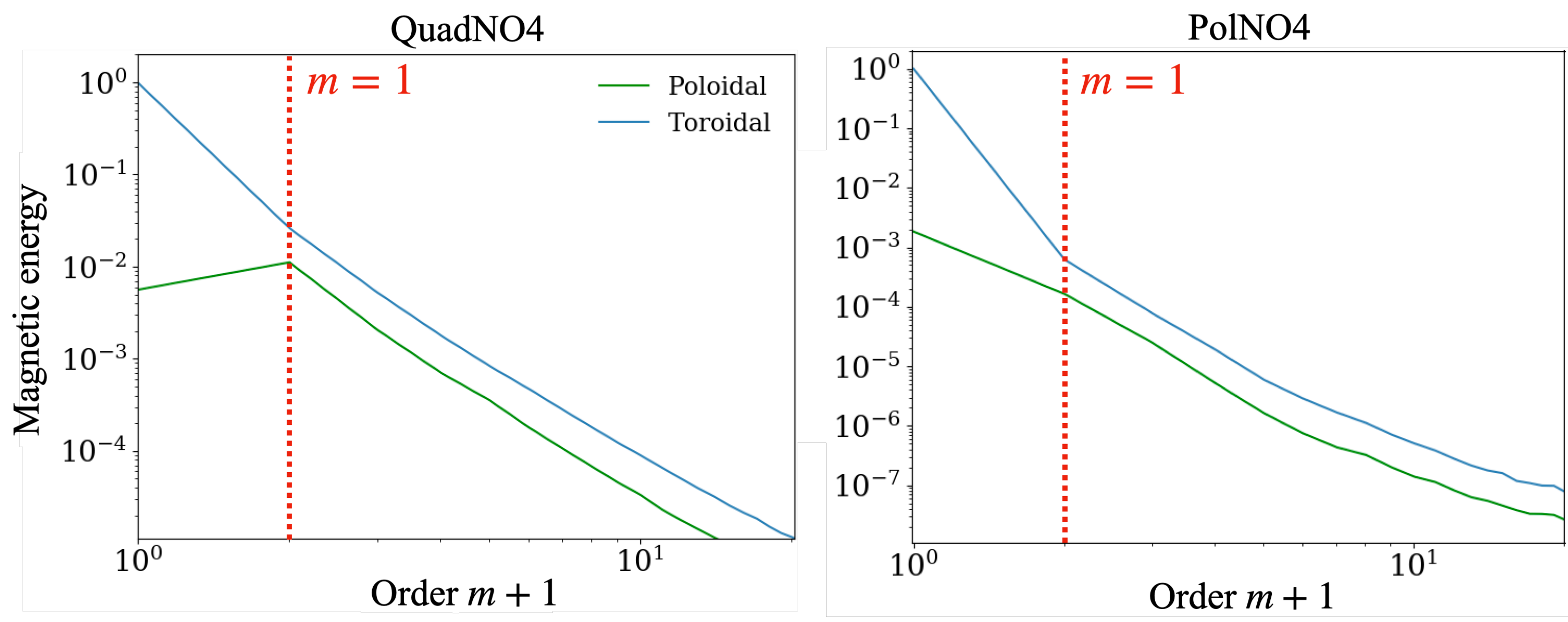}
    \caption{Time and volume-averaged $m$-spectra of the magnetic energy for both simulations on the equatorial and polar dynamos at $\overline{\Omega}/N=0.25,$  whose magnetic fields are displayed in Figs.~\ref{fig:bifurcation} and~\ref{fig:Ball}. The energy is rescaled by the energy of the most energetic mode: the $m=0$-toroidal component}
    \label{fig:spectra}
\end{figure*}

\section{Linear growth of the MHD instabilities driving the dynamos}\label{sec:tests}

\begin{figure*}[h!]
    \sidecaption
    \centering
    \includegraphics[width=12cm]{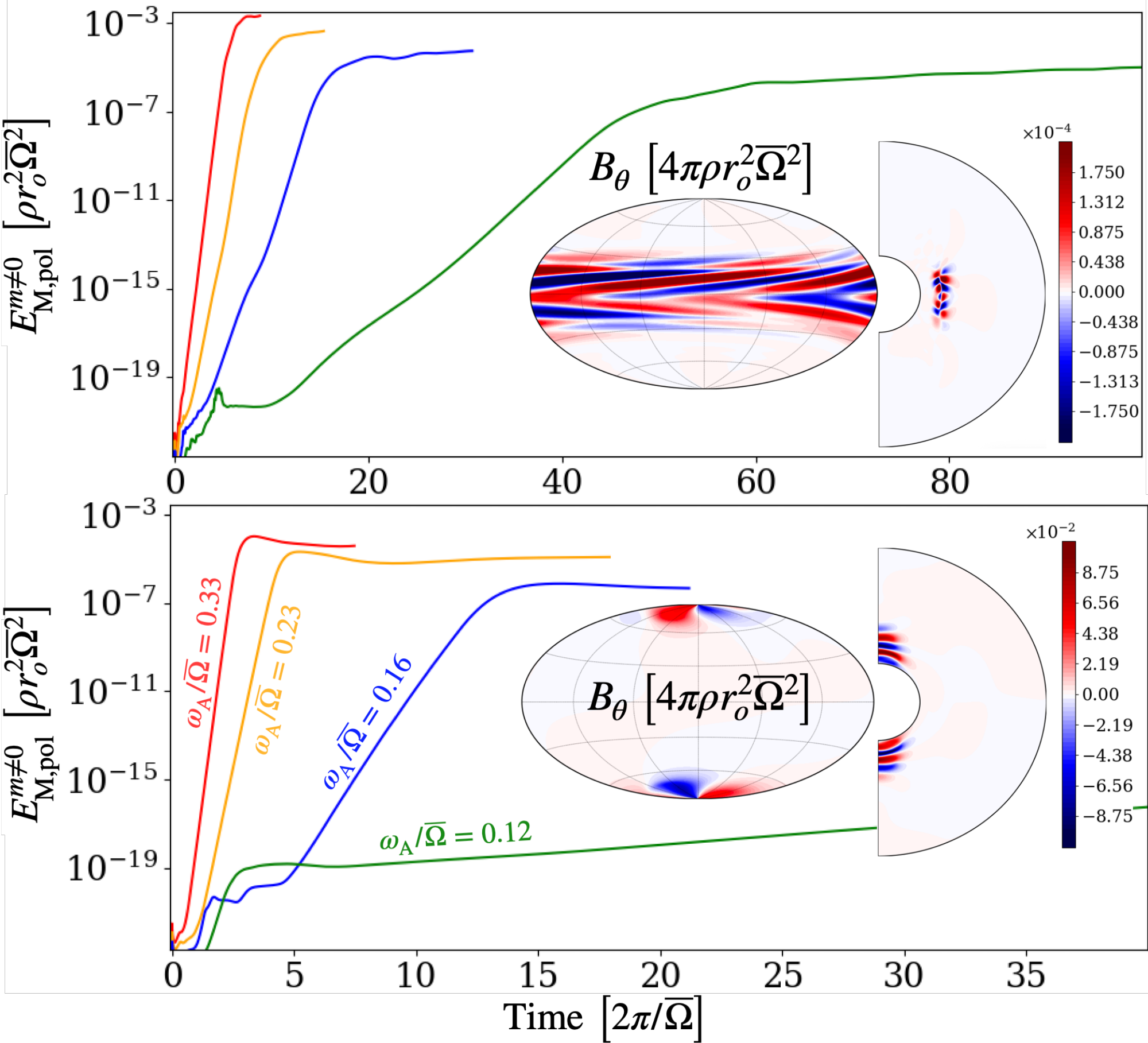}
    \caption{Time series of the energy associated with the non-axisymmetric poloidal component of the magnetic field for different values of $\omega_{\rm A}/\overline{\Omega}$. The runs represented on top include differential rotation ($q=-1$), while solid-body rotation is used for the runs in the bottom ($q=0$). The fixed input parameters are $\overline{\Omega}/N=0.25$, $Pm=4$, and $Pr=0.1$. The insets represent a hammer projection at the radius $r=0.35r_o$ (left) and a meridional slice of the latitudinal magnetic field $B_{\theta}$. For the simulations, insulating boundary conditions are used on both spheres.}
    \label{fig:Em_naxi}
\end{figure*}

To determine the nature of the MHD instabilities driving the different dynamos, we performed numerical simulations at $\overline{\Omega}/N=0.25$, where both dynamos coexist. While we initiate all runs with a $(\ell=1,m=0)$-azimuthal magnetic field, we include either differential rotation with $q=-1$ or a solid-body rotation ($q=0$). These simulations aim at investigating the linear growth of MHD instabilities that could develop in this setup and, notably, quantify the influence of the magnetic field strength and of rotation on the growth rate of these instabilities by varying the ratio $\omega_{\rm A}/\overline{\Omega}$.

In Fig.~\ref{fig:Em_naxi}, we observe that the growing MHD instability does not have the same location depending on the rotation radial profile: near the equator or near the polar axis when $q=-1$ (figure on the top) or $q=0$ (figure on the bottom). The location of the unstable modes indicates that the observed equatorial and polar MHD instabilities drive the equatorial and polar dynamos presented in this paper. We also observe that in the presence of differential rotation, the equatorial instability is the most unstable for a background $(\ell=1,m=0)$-azimuthal magnetic field, which is consistent with dynamo simulations and reinforces the link between the equatorial MHD instability and the dynamo.

\begin{figure*}[h!]
    \sidecaption
    \centering
    \includegraphics[width=12cm]{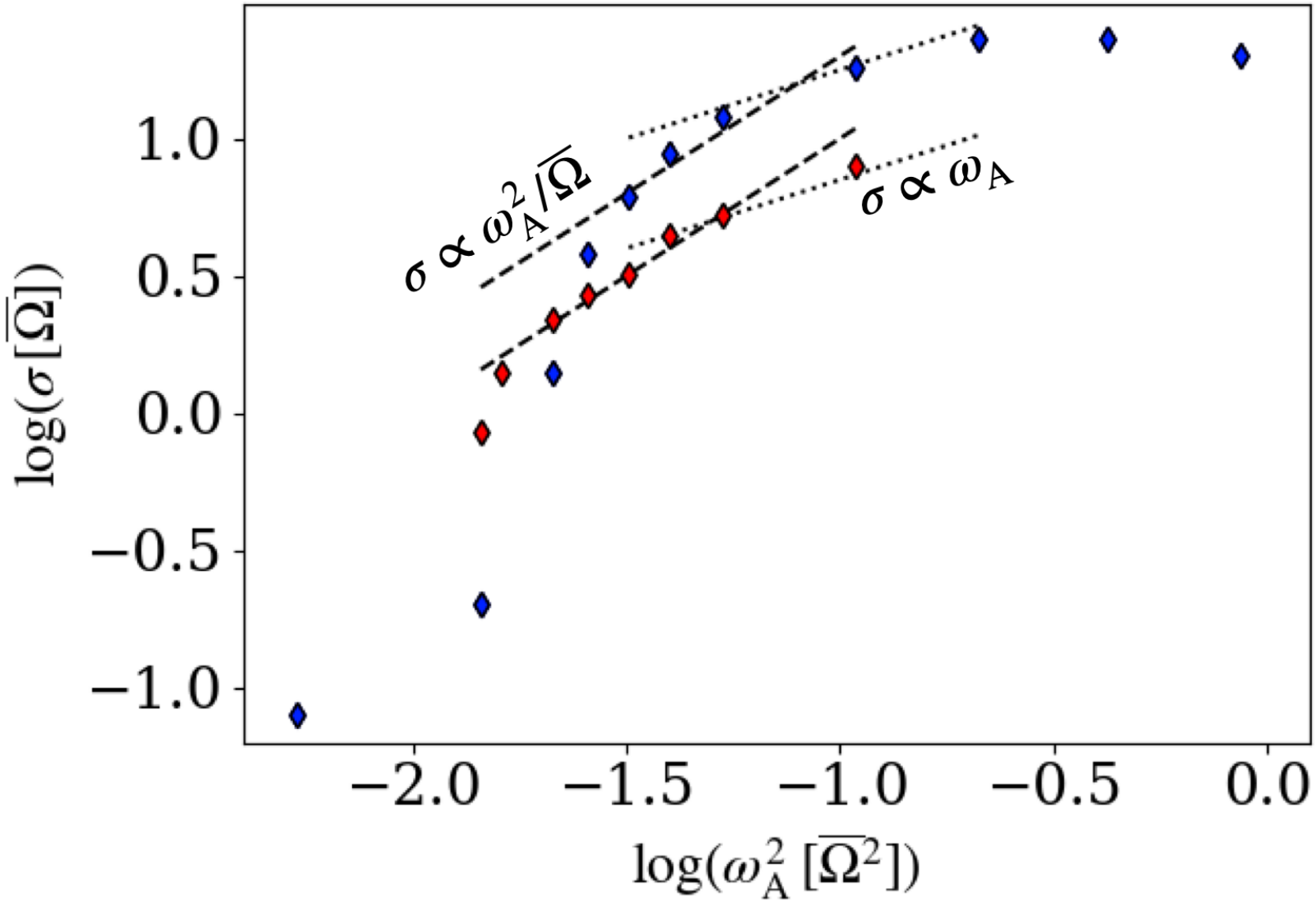}
    \caption{Growth rates of the energy associated with the non-axisymmetric poloidal component of the magnetic field as a function of the squared Alfvén frequency $\omega_{\rm A}^2$ (standard logarithmic representation). The flow is either in differential ($q=-1$, red) and solid-body ($q=0$, blue) rotation. Curves respecting $\sigma\propto\omega_{\rm A}$ (dotted lines) and $\sigma\propto\omega_{\rm A}^2/\overline{\Omega}$ (dashed lines) are overplotted. For the simulations, insulating boundary conditions are used on both spheres.}
    \label{fig:growth}
\end{figure*}

On the one hand, we observe that the polar instability has a kink structure because the only unstable modes are $m=1$. Moreover, the instability grows despite solid-body rotation, which implies that it must feed off the strong azimuthal magnetic field (and so vertical electric current). Therefore, the polar dynamo is certainly driven by the Tayler instability.

\begin{figure*}[h!]
    \sidecaption
    \centering
    \includegraphics[width=12cm]{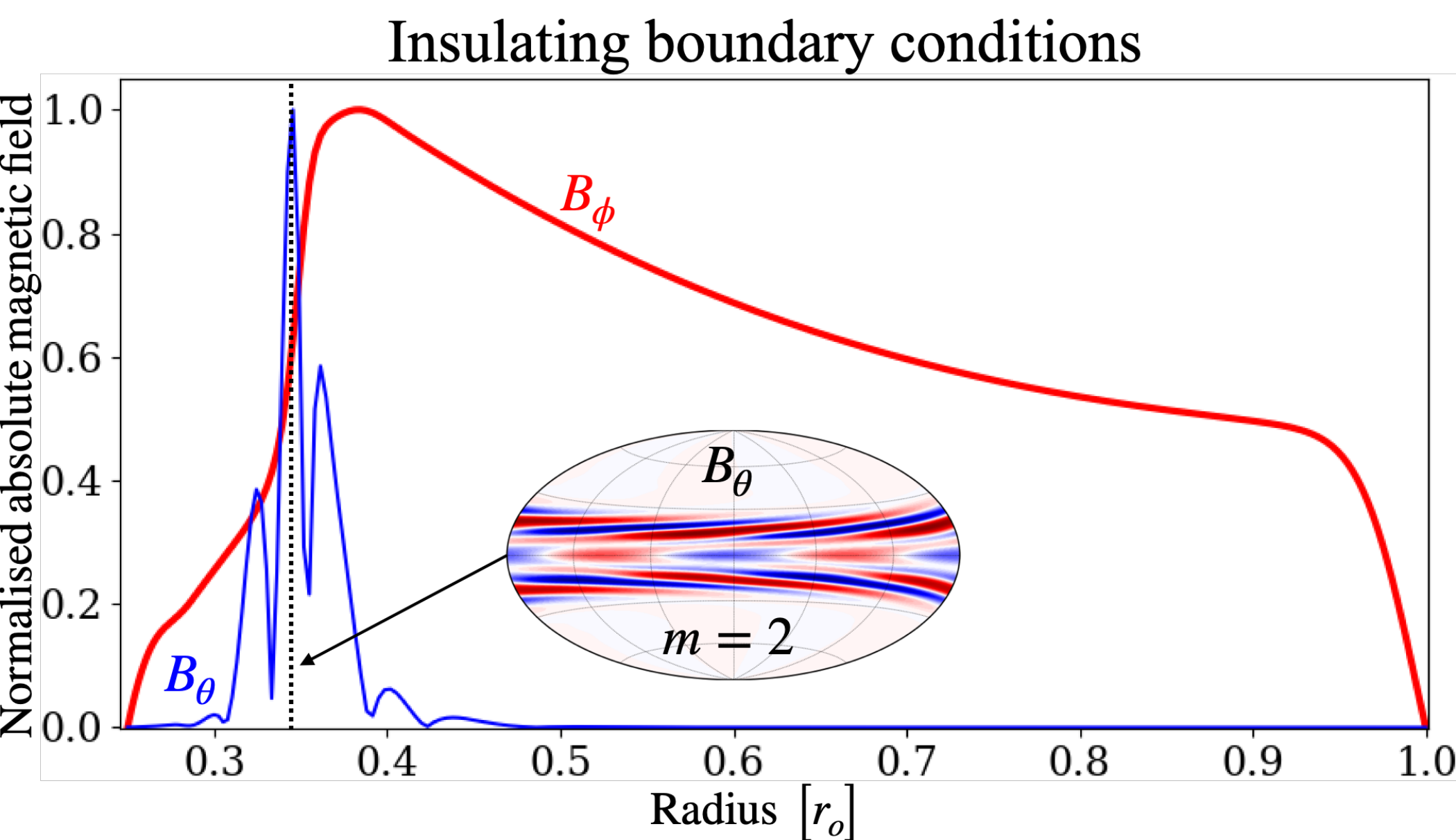}
    \caption{Radial profiles of the azimuthal ($B_{\phi}$, red curve) and latitudinal ($B_{\theta}$, blue curve) magnetic fields at the equator (absolute values normalised by the maximum). Hammer projections of $B_{\theta}$ at the radii indicated by a vertical dotted line. Insulating boundary conditions are used on both spheres.}
    \label{fig:profiles_IBC}
\end{figure*}

\begin{figure*}[h!]
    \sidecaption
    \centering
    \includegraphics[width=12cm]{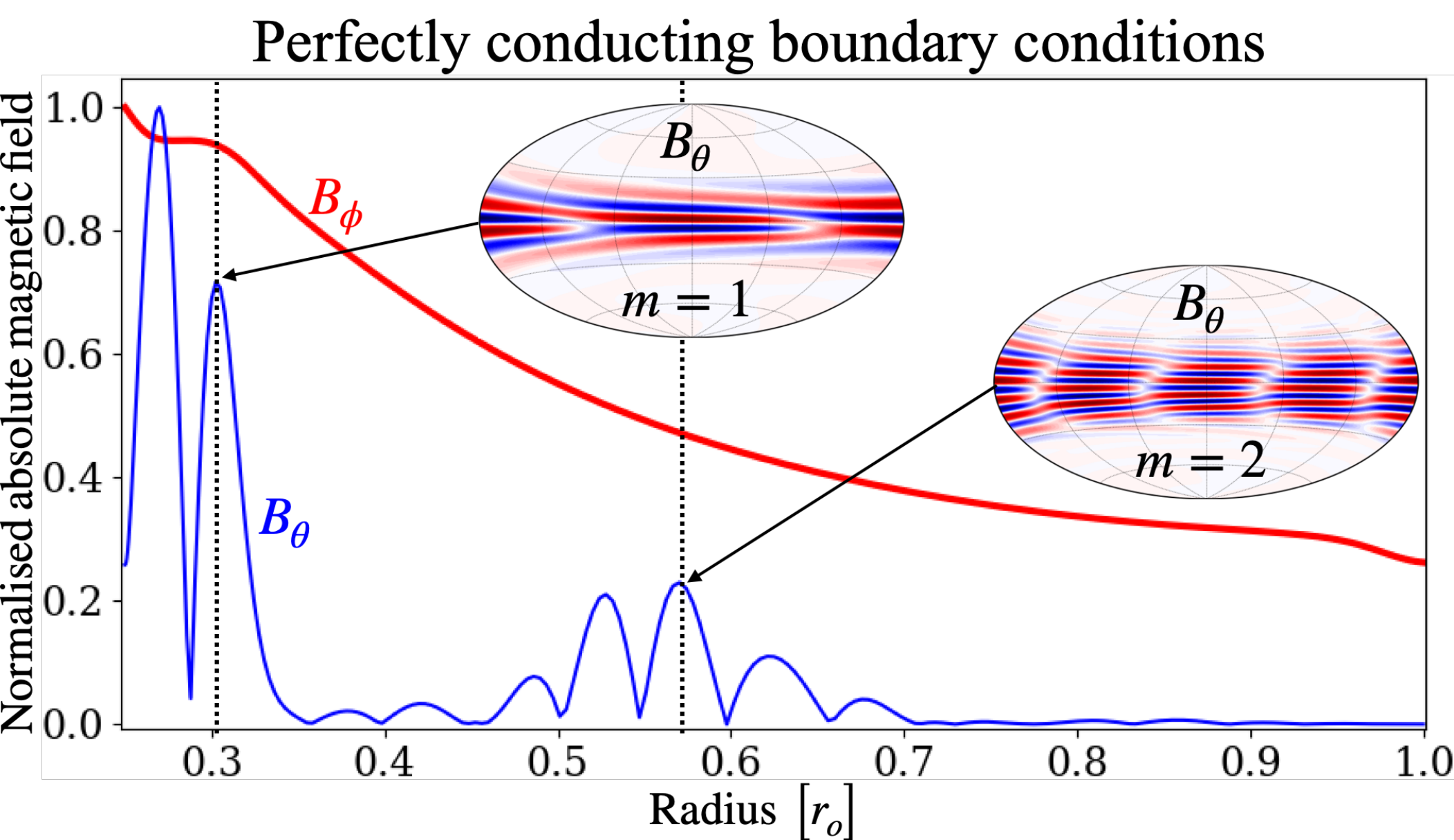}
    \caption{Same as Fig.~\ref{fig:profiles_IBC} but perfectly conducting boundary conditions.}
    \label{fig:profiles_PCBC}
\end{figure*}

On the other hand, the identification of the equatorial instability is more delicate. First, the instability appears in the presence of differential rotation, suggesting it feeds off shearing. Besides, the flow is not unstable to this instability in numerical simulations of a differentially rotating flow with a positive shear (q>0) and the same background azimuthal magnetic field. This explains why the equatorial dynamo has not been observed in the MHD simulations of~\citet{barrere2023,barrere2025}, who modelled a PNS interior whose surface rotates faster than the core. However, by measuring the slope in the linear phase of the time series of the energy associated with the poloidal non-axisymmetric magnetic field, we find that the growth rate of the instability (noted $\sigma$) follows similar power laws as the polar Tayler instability (see Fig.~\ref{fig:growth}): $\sigma\propto\omega_{\rm A}$ for $\omega_{\rm A}/\overline{\Omega}\in[0.23,0.46]$ and $\sigma\propto\omega_{\rm A}^2/\overline{\Omega}$ for $\omega_{\rm A}/\overline{\Omega}\in[0.15,0.2]$. These scalings correspond to the analytically predicted growth rates in the slow and fast rotating regimes, respectively~\citep{spruit1999}. Therefore, the instability is stabilised by rotation, unlike the standard MRI and grows similarly to the Tayler instability. Finally, focusing on the non-axisymmetric modes, we show in Fig.~\ref{fig:profiles_IBC} a snapshot of the radial profile of the latitudinal (blue) and azimuthal (red) magnetic fields ($B_{\theta}$ and $B_{\phi}$, respectively) at the equator. The most excited azimuthal mode is $m=2$ and develops where the radial gradient of $B_{\phi}$ is positive, while no modes are seen far from the boundary condition where $B_{\phi}\propto1/r$, i.e. the region is current-free. This suggests that the instability is also current-driven. Since these positive gradients of $B_{\phi}$ are due to the insulating boundary conditions, we reproduced the same simulation but with perfectly conducting boundary conditions to keep a cleaner radial profile $B_{\phi}\propto1/r$ at the equator from the inner to the outer boundary. The associated profiles are plotted in Fig.~\ref{fig:profiles_PCBC}. While $m=2$ modes transiently appear where $B_{\phi}\propto1/r$, signalling the presence of AMRI, $m=1$ modes grow and eventually dominate close to the inner boundary where the profile of $B_{\phi}$ is less steep. Note that the growth rate does not change with these new magnetic boundary conditions. This confirms the current-driven character of equatorial instability. Therefore, the instability shares both characteristics of the AMRI and the Tayler instability (TI). This type of instability has already been predicted in cylindrical Taylor-Couette flows by~\citet{kirillov2014} for negative shear. Our study confirms numerically the connection between the standard AMRI (differentially rotating and current-free flow) and Tayler instabilities (solid-body rotation and strong vertical currents). \citet{stefani2015} also predicts a link between AMRI and the Tayler instability for positive shear, but we only observe the classical Tayler instability in our simulation with positive shear.

\section{Measurement of the different quantities}\label{sec:average}
\begin{figure}[h!]
    \centering
    \includegraphics[width=\columnwidth]{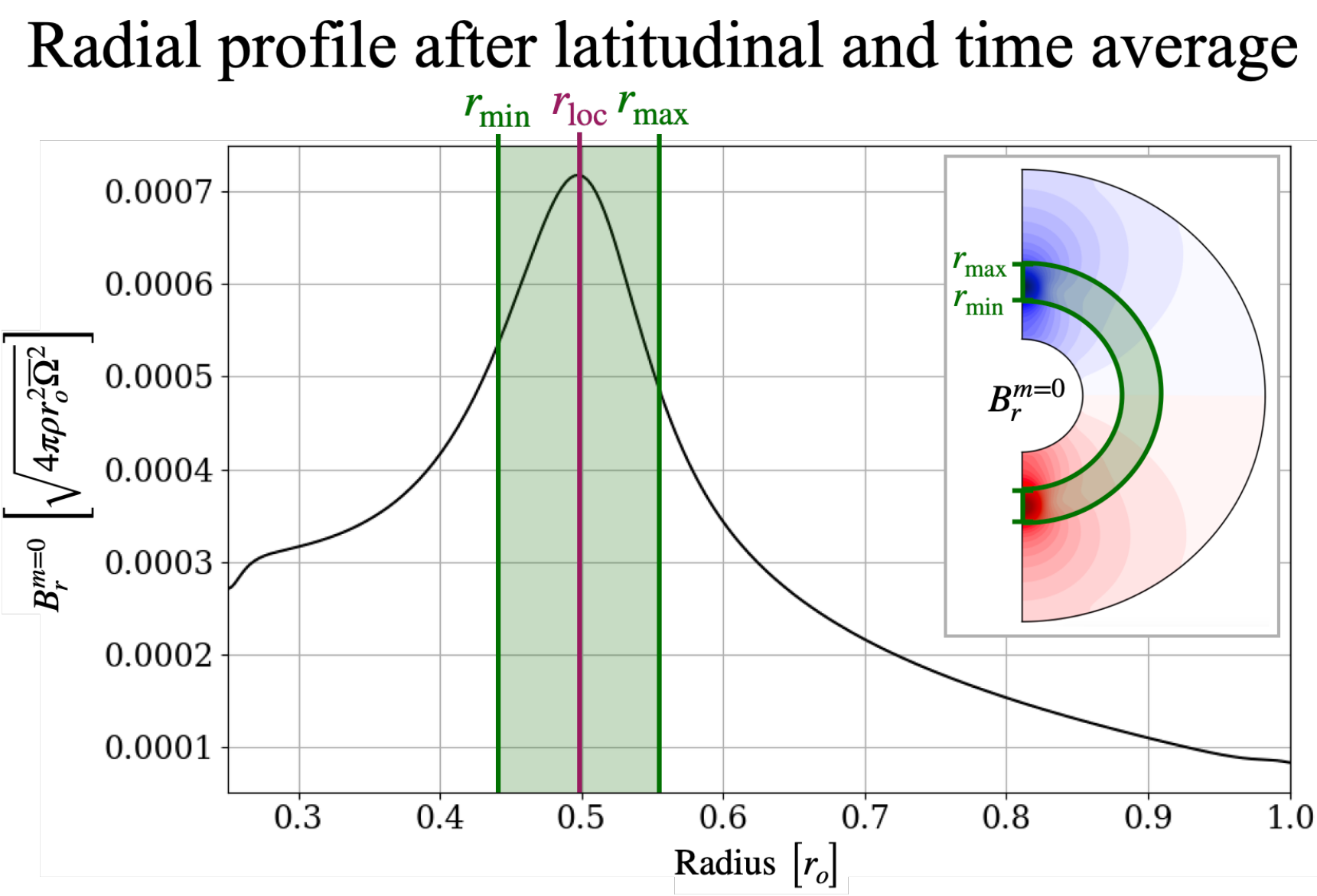}
    \caption{Radial profile of the axisymmetric radial magnetic field averaged in time and horizontally. The green zone represents the interval in radius over which we average a quantity. Inset: Meridional slice of the axisymmetric radial magnetic field on which the green zone is overplotted.}
    \label{fig:average}
\end{figure}

To measure the different field components, transports, the rotation rate, and the shear rate, we first produce a time- and latitudinally-averaged radial profile of these quantities (see Fig.~\ref{fig:average}). We then define an interval of radii $[r_{\min},r_{\rm max}]$ that corresponds to the range between which the energy associated to the axisymmetric radial magnetic field $E_{B_r}=(B_r^{m=0}/\sqrt{4\pi\rho})^2 > 0.5 {\,\rm max}(E_{B_r})$. $r_{\rm loc}$ is defined as the radius where $E_{B_r}$ is maximum. We finally estimate the quantities we use in the plots by calculating their root-mean-square of the quantities in $[r_{\min},r_{\rm max}]$.

\section{Local measure of $B_r/B_{\phi}$}\label{sec:BrBphi_loc}
\begin{figure}[h!]
    \centering
    \includegraphics[width=\columnwidth]{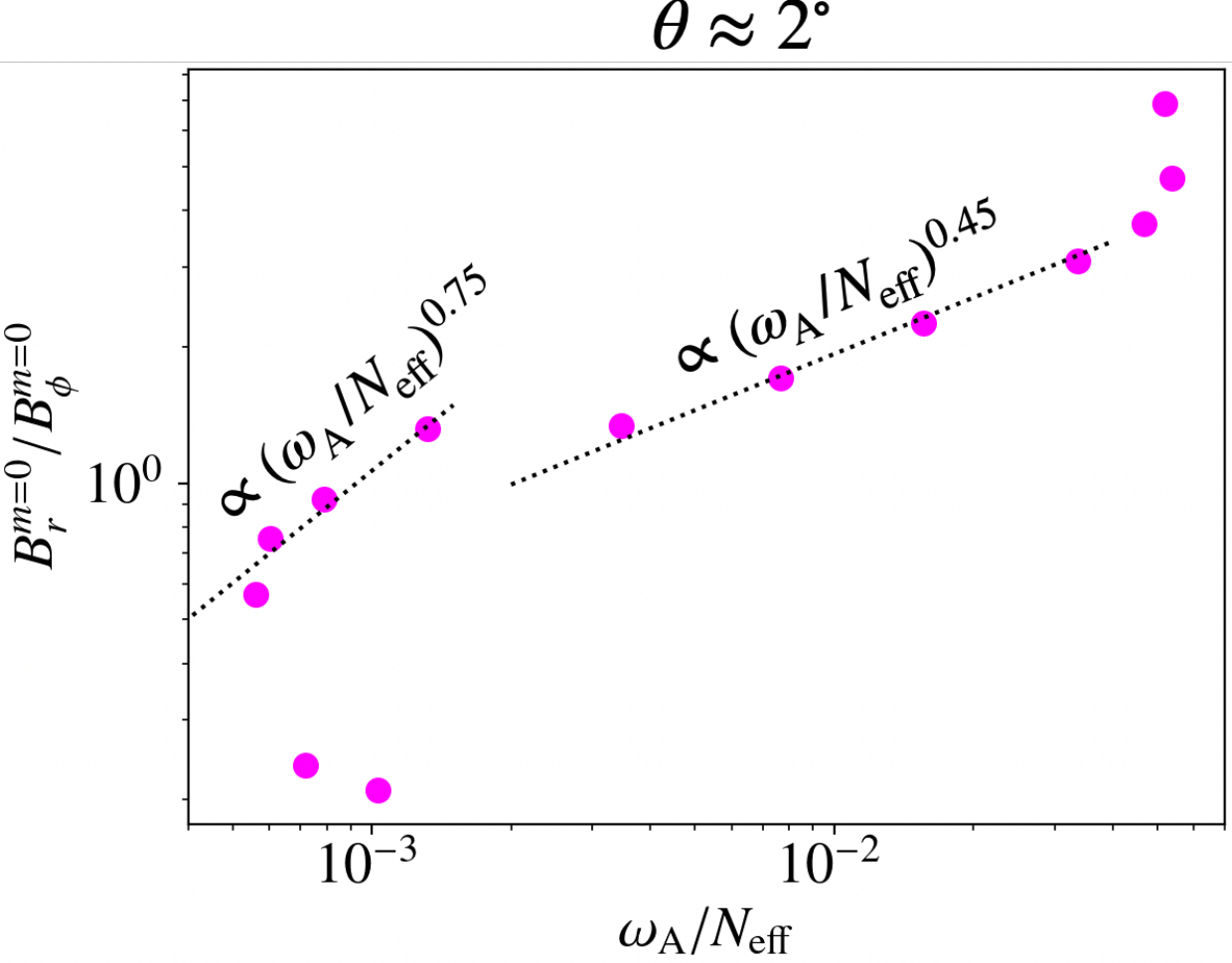}
    \caption{Ratio of the axisymmetric radial to the azimuthal magnetic field at the colatitudinal maximum of the radial component ($\theta\approx2^{\circ}$) as a function of the theoretical ratio of the Alfvén frequency to the effective Brunt-Väisälä frequency.}
    \label{fig:BrBp}
\end{figure}

In our simulations, the axisymmetric radial $B_r^{m=0}$ and azimuthal $B_{\phi}^{m=0}$ estimated as described in Sect:~\ref{sec:average} does not follows the relation in Eq.~\ref{eq:ratio_BrB}. To make sure it is not caused by our methods of estimating the strength of the magnetic field components, we also measured $B_r^{m=0}/B_{\phi}^{m=0}$ using a radial profile at the colatitude where $B_r^{m=0}$ is maximum ($\theta\approx2^{\circ}$), instead of averaging in latitude. Fig.~\ref{fig:BrBp} clearly shows that the relation in Eq.~\ref{eq:ratio_BrB} is not respected.

\section{List of models}\label{sec:models}
Tables~\ref{tab:sim_input} summarises the key parameters of the simulations carried out in this study, while Tables~\ref{tab:sim_loc}-\ref{tab:sim_nu} list the different quantities used to produce plots in the paper.

\begin{table*}[h!]
\caption{Input parameters}
\centering
\begin{tabular}{lcccccr}
\hline\hline
 Name & $E$ & $Ra$ & ${\overline{\Omega}/N}$ & ${\overline{\Omega}/N_{\rm eff}}$ & ${B_{\rm init}^{\rm rms}}$ & $(n_r,l_{\rm max},m_{\rm max})$\\
  &   &   &   &   & $\left[\sqrt{4\pi\rho r_o^2\overline{\Omega}^2}\right]$ &  \\
\hline
$ \text{QuadNO4} $ & $\SI{e-5}{}$ & $\SI{1.6e10}{}$ & $0.25$ & $1.6$ & $\sim 1$ & \text{$(256,170,170)$} \\
$ \text{QuadNO6} $ & $\SI{e-5}{}$ & $\SI{3.6e10}{}$ & $0.17$ & $1.1$ & \text{${B_{\rm init}^{\rm rms}}(\mathrm{QuadNO4})$} & \text{$(256,170,170)$} \\
$ \text{QuadNO8} $ & $\SI{e-5}{}$ & $\SI{6.4e10}{}$ & $0.13$ & $0.79$ & \text{${B_{\rm init}^{\rm rms}}(\mathrm{QuadNO6})$} & \text{$(256,170,170)$} \\
$ \text{PolNO2} $ & $\SI{e-5}{}$ & $\SI{4e9}{}$ & $0.5$ & $3.6$ & $\sim 1$ & \text{$(256,170,100)$} \\
$ \text{PolNO4} $ & $\SI{e-5}{}$ & $\SI{1.6e10}{}$ & $0.25$ & $1.6$ & $\sim 1$ & \text{$(256,170,100)$} \\
$ \text{PolNO6} $ & $\SI{e-5}{}$ & $\SI{3.6e10}{}$ & $0.17$ & $1.1$ & \text{${B_{\rm init}^{\rm rms}}(\mathrm{PolNO4})$} & \text{$(256,170,100)$} \\
$ \text{PolNO8} $ & $\SI{e-5}{}$ & $\SI{6.4e10}{}$ & $0.13$ & $0.79$ & \text{${B_{\rm init}^{\rm rms}}(\mathrm{PolNO6})$} & \text{$(256,170,100)$} \\
$ \text{PolNO10} $ & $\SI{e-5}{}$ & $\SI{e11}{}$ & $0.1$ & $0.63$ & \text{${B_{\rm init}^{\rm rms}}(\mathrm{PolNO8})$} & \text{$(256,170,100)$} \\
$ \text{PolNO15} $ & $\SI{e-5}{}$ & $\SI{2.25e11}{}$ & $0.07$ & $0.42$ & \text{${B_{\rm init}^{\rm rms}}(\mathrm{PolNO10})$} & \text{$(256,170,100)$} \\
$ \text{PolNO20} $ & $\SI{e-5}{}$ & $\SI{4e11}{}$ & $0.05$ & $0.32$ & \text{${B_{\rm init}^{\rm rms}}(\mathrm{PolNO15})$} & \text{$(320,256,63)$} \\
$ \text{PolNO30} $ & $\SI{1.5e-5}{}$ & $\SI{4e11}{}$ & $0.03$ & $0.21$ & \text{${B_{\rm init}^{\rm rms}}(\mathrm{PolNO20})$} & \text{$(320,256,63)$} \\
$ \text{PolNO50} $ & $\SI{2.5e-5}{}$ & $\SI{4e11}{}$ & $0.02$ & $0.13$ & \text{${B_{\rm init}^{\rm rms}}(\mathrm{PolNO30})$} & \text{$(320,256,63)$} \\
$ \text{PolNO70} $ & $\SI{3.5e-5}{}$ & $\SI{4e11}{}$ & $0.014$ & $0.09$ & \text{${B_{\rm init}^{\rm rms}}(\mathrm{PolNO50})$} & \text{$(320,256,63)$} \\
$ \text{PolNO100} $ & $\SI{5e-5}{}$ & $\SI{4e11}{}$ & $0.01$ & $0.063$ & \text{${B_{\rm init}^{\rm rms}}(\mathrm{PolNO70})$} & \text{$(320,256,63)$} \\
$ \text{PolNO110} $ & $\SI{5.5e-5}{}$ & $\SI{4e11}{}$ & $0.0091$ & $0.057$ & \text{${B_{\rm init}^{\rm rms}}(\mathrm{PolNO100})$} & \text{$(320,128,29)$} \\
$ \text{PolNO130} $ & $\SI{6.5e-5}{}$ & $\SI{4e11}{}$ & $0.0077$ & $0.049$ & \text{${B_{\rm init}^{\rm rms}}(\mathrm{PolNO110})$} & \text{$(320,170,49)$} \\
\hline
\end{tabular}
\tablefoot{Overview of the input parameters used for the different simulations. All the simulations have the same aspect ratio $\chi=0.25$, thermal and magnetic Prandtl numbers $Pr=0.1$ and $Pm=4$.}
\label{tab:sim_input}
\end{table*}

\begin{table*}[h!]
 \caption{Different quantities measured in the 3D simulations}
 \centering
\begin{tabular}{lcccccccccr}
\hline\hline
 Name & $E^{m\neq0}_{\rm M}$ & $l_{\rm TI}$ & $2\eta \Omega_{\rm loc}/\omega_{\rm A}^2$ & $2\omega_{\rm A}/N_{\rm eff}$ & $a$ & $r_{\rm min}$ & $r_{\rm max}$ & $r_{\rm loc}$ & $\Omega_{\rm loc}$ & $q$ \\
  & $\left[\rho r_o^2\overline{\Omega}^2\right]$ & $[r_o]$ & $[r_o]$ & $[r_o]$ &  & $[r_o]$ & $[r_o]$ & $[r_o]$ & $\left[\overline{\Omega}\right]$ & \\
\hline
$ \text{QuadNO4} $ & $\SI{5.2e-5}{}$ & -- & -- & -- & $0.4$ & -- & -- & -- & -- & -- \\
$ \text{QuadNO6} $ & $\SI{2.2e-5}{}$ & -- & -- & -- & $0.59$ & -- & -- & -- & -- & -- \\
$ \text{QuadNO8} $ & $\SI{3.8e-6}{}$ & -- & -- & -- & -- & -- & -- & -- & -- & -- \\
$ \text{PolNO2} $ & $\SI{1.3e-5}{}$ & $0.22$ & $0.047$ & $0.45$ & $0.88$ & $0.29$ & $0.60$ & $0.35$ & $1.1$ & $-0.11$ \\
$ \text{PolNO4} $ & $\SI{3.5e-7}{}$ & $0.11$ & $0.037$ & $0.28$ & $0.98$ & $0.31$ & $0.62$ & $0.47$ & $1.1$ & $-0.12$ \\
$ \text{PolNO6} $ & $\SI{1.1e-7}{}$ & $0.09$ & $0.034$ & $0.20$ & $0.97$ & $0.42$ & $0.66$ & $0.54$ & $1.1$ & $-0.13$ \\
$ \text{PolNO8} $ & $\SI{1.2e-7}{}$ & $0.075$ & $0.036$ & $0.14$ & $0.97$ & $0.46$ & $0.67$ & $0.57$ & $1.0$ & $-0.13$ \\
$ \text{PolNO10} $ & $\SI{1.1e-7}{}$ & $0.058$ & $0.037$ & $0.11$ & $0.97$ & $0.53$ & $0.71$ & $0.63$ & $1.0$ & $-0.13$ \\
$ \text{PolNO15} $ & $\SI{1.2e-7}{}$ & $0.056$ & $0.035$ & $0.077$ & $0.95$ & $0.59$ & $0.71$ & $0.66$ & $1.0$ & $-0.13$ \\
$ \text{PolNO20} $ & $\SI{6.6e-8}{}$ & $0.046$ & $0.036$ & $0.056$ & $0.91$ & $0.66$ & $0.71$ & $0.69$ & $1.0$ & $-0.13$ \\
$ \text{PolNO30} $ & $\SI{6.2e-8}{}$ & $0.051$ & $0.031$ & $0.054$ & $0.88$ & $0.55$ & $0.66$ & $0.60$ & $1.1$ & $-0.20$ \\
$ \text{PolNO50} $ & $\SI{1.1e-7}{}$ & $0.046$ & $0.031$ & $0.042$ & $0.87$ & $0.54$ & $0.63$ & $0.59$ & $1.1$ & $-0.31$ \\
$ \text{PolNO70} $ & $\SI{1.1e-7}{}$ & $0.034$ & $0.032$ & $0.037$ & $0.85$ & $0.44$ & $0.65$ & $0.58$ & $1.2$ & $-0.42$ \\
$ \text{PolNO100} $ & $\SI{4.8e-8}{}$ & $0.034$ & $0.028$ & $0.037$ & $0.80$ & $0.44$ & $0.58$ & $0.52$ & $1.3$ & $-0.57$ \\
$ \text{PolNO110} $ & $\SI{2.7e-8}{}$ & $0.032$ & $0.027$ & $0.037$ & $0.74$  & $0.44$ & $0.56$ & $0.51$ & $1.3$ & $-0.62$\\
$ \text{PolNO130} $ & $\SI{2.8e-7}{}$ & $0.03$ & $0.026$ & $0.036$ & $0.72$ & $0.44$ & $0.55$ & $0.50$ & $1.4$ & $-0.71$ \\
\hline
\end{tabular}
\tablefoot{Values of the volume and time-averaged turbulent magnetic energy ($E^{m\neq 0}_{\rm turb}$), the radial length-scale of the Tayler instability ($l_{\rm TI}$) and its theoretical bottom ($2\eta\Omega_{\rm loc}/\omega_{\rm A}^2$) and top ($2\omega_{\rm A}/N_{\rm eff}$) limits, and the asymmetry parameter ($a$) displayed in Figs.~\ref{fig:bifurcation},~\ref{fig:lTI}, and~\ref{fig:asym}, respectively. The values of the local radii ($r_{\rm min},$ $r_{\rm max},$ and $r_{\rm loc}$), rotation rate ($\Omega_{\rm loc}$), and the shear rate ($q$) --- used to average the different quantities to estimate the scaling laws --- are also listed.}
\label{tab:sim_loc}
\end{table*}

\begin{table*}[h!]
\caption{Averaged magnetic fields}
\centering
\begin{tabular}{lccccr}
\hline\hline
 Name & $B_{\phi}^{m=0}$ & $B_{r}^{m=0}$ & $B_{\rm tot}^{m\neq0}$ & $B_{\perp}^{m\neq0}$ & $B_{r}^{m\neq0}$ \\
  & $\left[10^{-3}\sqrt{4\pi\rho r_{\rm loc}^2\Omega_{\rm loc}^2}\right]$ & $\left[10^{-3}\sqrt{4\pi\rho r_{\rm loc}^2\Omega_{\rm loc}^2}\right]$ & $\left[10^{-3}\sqrt{4\pi\rho r_{\rm loc}^2\Omega_{\rm loc}^2}\right]$ & $\left[10^{-3}\sqrt{4\pi\rho r_{\rm loc}^2\Omega_{\rm loc}^2}\right]$ & $\left[10^{-3}\sqrt{4\pi\rho r_{\rm loc}^2\Omega_{\rm loc}^2}\right]$ \\
\hline
$ \text{PolNO2} $ & $65$ & $30$ & $21$ & $19$ & $8.6$ \\
$ \text{PolNO4} $ & $82$ & $29$ & $5.0$ & $4.3$ & $2.4$ \\
$ \text{PolNO6} $ & $90$ & $25$ & $5.1$ & $4.4$ & $2.5$ \\
$ \text{PolNO8} $ & $86$ & $18$ & $4.6$ & $4.1$ & $2.0$ \\
$ \text{PolNO10} $ & $84$ & $8.0$ & $3.2$ & $2.9$ & $1.1$ \\
$ \text{PolNO15} $ & $89$ & $5.1$ & $2.9$ & $2.7$ & $0.86$ \\
$ \text{PolNO20} $ & $86$ & $2.8$ & $2.7$ & $2.6$ & $0.78$ \\
$ \text{PolNO30} $ & $120$ & $1.7$ & $3.1$ & $3.0$ & $0.51$ \\
$ \text{PolNO50} $ & $150$ & $1.6$ & $3.6$ & $3.6$ & $0.29$ \\
$ \text{PolNO70} $ & $170$ & $1.3$ & $2.7$ & $2.7$ & $0.11$ \\
$ \text{PolNO100} $ & $220$ & $1.3$ & $2.8$ & $2.8$ & $0.072$ \\
$ \text{PolNO110} $ & $240$ & $0.98$ & $2.7$ & $2.7$ & $0.063$ \\
$ \text{PolNO130} $ & $260$ & $0.89$ & $2.7$ & $2.7$ & $0.059$ \\
\hline
\end{tabular}
\tablefoot{Different magnetic field components of Fig.~\ref{fig:Ball}: axisymmetric azimuthal ($B_{\phi}^{m=0}$), axisymmetric radial ($B_{r}^{m=0}$), total non-axisymmetric ($B_{\rm tot}^{m\neq0}$), non-axisymmetric perpendicular/horizontal ($B_{\perp}^{m=0}$), and non-axisymmetric radial ($B_{r}^{m\neq0}$).}
\label{tab:sim_Bloc}
\end{table*}

\begin{table}[h!]
\centering
\caption{Averaged viscosities associated with stress tensors}
\begin{tabular}{lccr}
\hline\hline
 Name & $\nu_{\rm M}$ & $\nu_{\rm R}$ & $\nu_{\rm mix}$ \\
  & $\left[r_{\rm loc}^2\Omega_{\rm loc}\right]$ & $\left[r_{\rm loc}^2\Omega_{\rm loc}\right]$ & $\left[r_{\rm loc}^2\Omega_{\rm loc}\right]$ \\
\hline
$ \text{PolNO2} $ & $\SI{5.9e-3}{}$ & $\SI{6.5e-6}{}$ & $\SI{2.1e-6}{}$ \\
$ \text{PolNO4} $ & $\SI{4.9e-3}{}$ & $\SI{1.0e-6}{}$ & $\SI{3.9e-7}{}$ \\
$ \text{PolNO6} $ & $\SI{3.4e-3}{}$ & $\SI{8.1e-7}{}$ & $\SI{5.4e-7}{}$ \\
$ \text{PolNO8} $ & $\SI{2.1e-3}{}$ & $\SI{4.5e-7}{}$ & $\SI{2.1e-7}{}$ \\
$ \text{PolNO10} $ & $\SI{6.1e-4}{}$ & $\SI{2.0e-7}{}$ & $\SI{7.0e-8}{}$ \\
$ \text{PolNO15} $ & $\SI{3.5e-4}{}$ & $\SI{1.3e-7}{}$ & $\SI{4.8e-8}{}$ \\
$ \text{PolNO20} $ & $\SI{1.7e-4}{}$ & $\SI{6.0e-8}{}$ & $\SI{4.3e-8}{}$ \\
$ \text{PolNO30} $ & $\SI{1.4e-4}{}$ & $\SI{6.3e-8}{}$ & $\SI{1.2e-7}{}$ \\
$ \text{PolNO50} $ & $\SI{1.3e-4}{}$ & $\SI{5.1e-8}{}$ & $\SI{4.0e-8}{}$ \\
$ \text{PolNO70} $ & $\SI{1.2e-4}{}$ & $\SI{4.6e-8}{}$ & $\SI{7.0e-8}{}$ \\
$ \text{PolNO100} $ & $\SI{1.2e-4}{}$ & $\SI{4.5e-8}{}$ & $\SI{6.1e-8}{}$ \\
$ \text{PolNO110} $ & $\SI{1.0e-4}{}$ & $\SI{3.0e-8}{}$ & $\SI{3.9e-8}{}$ \\
$ \text{PolNO130} $ & $\SI{9.3e-5}{}$ & $\SI{2.2e-8}{}$ & $\SI{5.7e-8}{}$ \\
\hline
\end{tabular}
\tablefoot{Table presenting the different viscosities of Fig.~\ref{fig:AMT} associated to the Maxwell stress ($\nu_{\rm M}$), the Reynolds stress ($\nu_{\rm R}$), and the chemical mixing ($\nu_{\rm mix}$).}
\label{tab:sim_nu}
\end{table}

\end{appendix}
\end{document}